%% file: main.tex
\documentclass{article}

\usepackage[preprint]{neurips_2026}
\input{00.packages}
\input{00.macros}

\input{00.metadata}

\begin{document}

\maketitle


\begingroup
\renewcommand{\thefootnote}{*}
\footnotetext{Equal Contribution. \hspace{1em} \textbf{Correspondence to:} Dhairya Kuchhal (\href{mailto:cs5240396@cse.iitd.ac.in}{\texttt{cs5240396@cse.iitd.ac.in}})}
\renewcommand{\thefootnote}{\dag}
\footnotetext{Equal Contribution.}
\renewcommand{\thefootnote}{\ddag}
\footnotetext{Joint Supervision.}
\setcounter{footnote}{0} 
\endgroup


%
\input{sections/00.abstract}

\input{sections/01.introduction}
\input{sections/02.data_curation}
\input{sections/03.benchmark_design}
\input{sections/04.main_results}
\input{sections/05.ablations}

\input{sections/06.conclusion}
\bibliographystyle{plainnat}
\bibliography{references}

\appendix
\input{sections/appendix_data_curation}
\input{sections/appendix_interpretations}
\input{sections/appendix_data_scaling}
\input{sections/appendix_transfer}
\input{sections/appendix_technical_report}
\end{document}

%% file: 00.packages.tex
\usepackage[utf8]{inputenc}
\usepackage[T1]{fontenc}
\usepackage{hyperref}
\usepackage{url}
\usepackage{booktabs}
\usepackage{authblk}
\usepackage{amsfonts}
\usepackage{amsmath}
\usepackage{amssymb}
\usepackage{nicefrac}
\usepackage{microtype}
\usepackage[table]{xcolor}
\usepackage{graphicx}
\graphicspath{{../}}
\usepackage{subcaption}
\usepackage{multirow}
\usepackage{array}
\usepackage{tabularx}
\usepackage{enumitem}
\usepackage{xspace}

\hypersetup{
  colorlinks=true,
  linkcolor=blue!60!black,
  citecolor=blue!60!black,
  urlcolor=blue!70!black,
}

%% file: 00.macros.tex
\newcommand{\scthree}{SC\textsuperscript{3}\xspace}

\newcommand{\logS}{\ensuremath{\log_{10} S}\xspace}
\newcommand{\logSunit}{\ensuremath{\log_{10}(\text{mol\,L}^{-1})}\xspace}

\newcommand{\epsA}{\ensuremath{\varepsilon_{\text{aleatoric}}}\xspace}

\newcommand{\RMSE}{\textsc{rmse}\xspace}
\newcommand{\Rr}{\ensuremath{R^2}\xspace}

\newcommand{\eg}{\textit{e.g.}\xspace}

\newcommand{\old}[1]{#1}
\newcommand{\numval}[1]{\textcolor{blue!70!black}{#1}}

%% file: 00.metadata.tex
\title{SC\textsuperscript{3}: The Multi-Solvent Solubility Challenge and Benchmark}

\author[1]{Vansh Ramani}
\author[1]{Har Ashish Arora\textsuperscript{*}}
\author[1]{Dhairya Kuchhal\textsuperscript{*}}
\author[2]{Sergei Tatarin\textsuperscript{\dag}}
\author[2]{Lev Krasnov\textsuperscript{\dag}}
\author[1]{Sayan Ranu\textsuperscript{\ddag}}
\author[1]{Tarak Karmakar\textsuperscript{\ddag}}

\affil[1]{Indian Institute of Technology Delhi, India}
\affil[2]{Kurnakov Institute of General and Inorganic Chemistry RAS, Russia}

%% file: sections/00.abstract.tex
\begin{abstract}
Solubility prediction is a standard benchmark in computational
chemistry, yet multi-solvent models which reportedly approach the
experimental-noise ceiling (\textit{i.e.} the \textit{aleatoric limit}) are not yet reliable enough to be deployed.  We argue that this gap is partly artefactual: published
benchmarks differ in curation policies, evaluate on count-weighted
RMSE that hides failure on tail-heavy solvent distributions, and treat the widely cited $0.6$--$0.8\,\log S$ inter-laboratory figure as the aleatoric ceiling even though it reflects worst-case, not expected, disagreement. We introduce \scthree, a multi-solvent solubility benchmark built on
\textsc{BigSolDB}~v2.1 with three contributions:
(i)~a reproducible curation pipeline yielding {101\,535}
measurements over {1\,327} solutes and {206} solvents,
with a recalibrated aleatoric floor of $0.106\,\log
S$---roughly~$6\times$ tighter than the conventional figure;
(ii)~nested Gold/Silver/Bronze consensus tiers with per-point
$\sigma$, three leakage-checked splits, and a multi-solvent metric
suite (PS-RMSE, $Z$-RMSE); and
(iii)~a {31}-model benchmark across six families, whose best
Bronze PS-RMSE sits at
$\approx\!5\times\varepsilon_{\mathrm{aleatoric}}$, and we observe this is a gap unclosed
by any deep alternative tested.
We perform three follow-on analyses: data scaling, transfer from
quantum-chemistry solvation energies, and feature-level
attribution, which demonstrates that calibrated per-point uncertainty is a reusable infrastructure for diagnosis beyond point prediction.
\end{abstract}

%% file: sections/01.introduction.tex

\section{Introduction}
\label{sec:introduction}

Solubility---the maximum equilibrium concentration of a solute in a
given solvent at fixed temperature and pressure---is one of the most
consequential and most measured properties in chemistry. It governs bioavailability and formulation in drug
discovery~\citep{ma2025solecos,bolla2022crystals}, dictates solvent
choice in flow and batch synthesis~\citep{reichardt2011solvents,diorazio2016toward,tu2025askcos},
controls polymorph selection in crystallisation~\citep{sheikholeslamzadeh2012optimal,mendis2022simultaneous},
and sets the environmental fate of intermediates. Experimental measurements are slow and expensive: a single
solute--solvent--temperature point can take days of time. With the chemical space relevant to drug discovery alone running into billions
of candidate solute--solvent pairs, a predictive model that
generalises across that space---telling a chemist what to make and
where to dissolve it before any glassware is touched---would compress
weeks of wait into seconds and is one of the longest-standing
open problems in cheminformatics. The community has accordingly
invested in scale: \textsc{AqSolDB}~\citep{sorkun2019aqsoldb},
\textsc{BigSolDB}~\citep{krasnov2025bigsoldb}, and
\textsc{MixtureSolDB}~\citep{malikov2026dataset} together aggregate
hundreds of thousands of measurements, and recent neural and
foundational models report errors approaching the inter-laboratory
floor~\citep{boobier2020machine,attia2025data,al2025accurately,jung2025enhancing,fowles2025physics}.

Yet deployment lags reported accuracy by a wide margin, and the
disconnect is largely how the field measures itself. \textit{Three
problems compound.}
\textit{(i) Inconsistent curation.}
Published splits over the same source databases apply different unit
conventions, duplicate-handling rules, and stereochemistry policies,
so reported numbers do not transfer between
studies~\citep{llompart2024curation}.
\textit{(ii) Single-axis evaluation.}
Aggregate RMSE is dominated by high-frequency solvents, rewarding memorisation of per-solvent location shifts rather than chemistry; failure on the long-tail solvents that matter most for novel formulation is invisible in the headline number.
\textit{(iii) A mis-calibrated aleatoric floor.}
The widely cited $0.6$--$0.8\,\log S$ inter-lab
figure~\citep{palmer2014dataquality} does \textit{not} reflect expected measurement noise; treating it as the noise ceiling
concedes an order of magnitude of measurable signal. We show in \S\ref{sec:aleatoric}
that the expected inter-lab disagreement is roughly $6\times$ smaller.

We introduce \scthree, a multi-solvent solubility benchmark designed
to address these issues. Our contributions are:
\begin{itemize}[leftmargin=1.2em,itemsep=2pt,topsep=2pt]
  \item \textbf{Reproducible curation with a recalibrated aleatoric
    floor.}  A stereo-preserving canonicalisation, two-stage
    duplicate-DOI detection, and seven-stage cleaning waterfall yield
    {101\,535} measurements over {1\,327} solutes and
    {206} solvents.  From {481} multi-source pairs we
    estimate
    $\varepsilon_{\mathrm{aleatoric}} = 0.106\,\log S$---roughly
    $6\times$ tighter than the conventional 0.6--0.8 figure
    (\S\ref{sec:aleatoric}).

  \item \textbf{A standardised multi-solvent benchmarking protocol.}
    Nested consensus tiers (Gold $\subset$ Silver $\subset$ Bronze)
    with per-point~$\sigma$; three leakage-checked splits
    (\textsc{Eval}, \textsc{OOD}, tiers); and a metric suite headlined
    by per-solvent RMSE (PS-RMSE) and noise-normalised $Z$-RMSE
    (\S\ref{sec:benchmark}--\ref{sec:metrics}).

  \item \textbf{A 31-model benchmark.}  We evaluate six model families (thermodynamic,
    descriptor + tree, fingerprint, descriptor-deep, graph, and
    foundation) under fixed splits and seeds (\S\ref{sec:baselines}) on \scthree.

  \item \textbf{Three analyses enabled by calibrated uncertainty.}
    We fit power-law scaling curves whose asymptotes lie above
    the aleatoric floor, showing that the residual error is due to a representation bottleneck and not a lack of data
    (\S\ref{sec:data_scaling}). Pretraining on
    ${\sim}10^{6}$ quantum-chemistry solvation
    energies~\citep{vermeire2021transfer} partially closes this gap,
    confirming that adjacent-task structure transfers to solubility
    (\S\ref{sec:transfer}).  Finally, SHAP and occlusion attribution
    reveal that LightGBM independently recovers the General Solubility
    Equation axes while a GCN learns a BRICS substructure ontology
    consistent with medicinal-chemistry intuition
    (\S\ref{sec:interpretability}).
\end{itemize}

All code, splits, curve fits, training scripts, and pretrained checkpoints accompanying the paper can be found at \href{https://anonymous.4open.science/r/SC3-Benchmark}{\texttt{https://anonymous.4open.science/r/SC3-Benchmark}}.

%% file: sections/02.data_curation.tex
\section{Data curation and aleatoric limit}
\label{sec:curation_aleatoric}

\subsection{Data curation pipeline}
\label{sec:raw_audit}

\begin{figure}[t]
\centering
\begin{subfigure}[t]{0.48\linewidth}
  \centering
  \includegraphics[width=\linewidth]{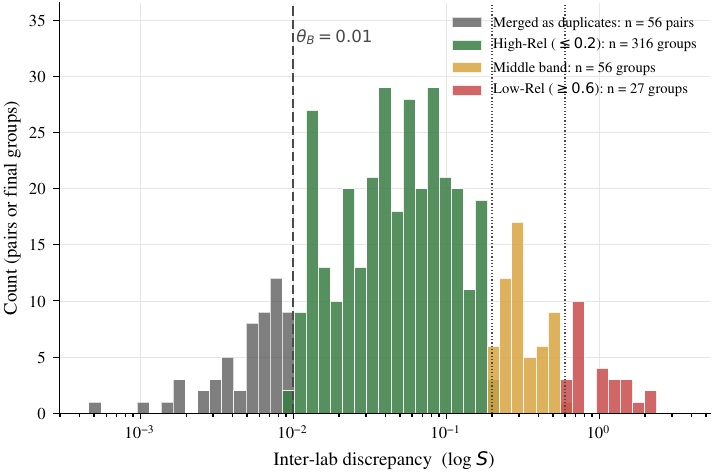}
  \caption{Source integrity filtering and ranking.}
  \label{fig:stages_bc}
\end{subfigure}%
\hfill
\begin{subfigure}[t]{0.48\linewidth}
  \centering
  \includegraphics[width=\linewidth]{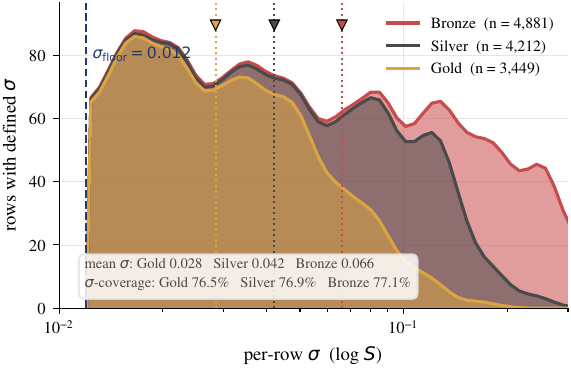}
  \caption{Per-row $\sigma$ distribution nested tiers (log-$\sigma$ axis).}
  \label{fig:tier_sigma}
\end{subfigure}
\caption{Data quality infrastructure. \textbf{(a)} Inter-lab discrepancy distribution driving the source-integrity pipeline (\S\ref{app:source_integrity}).  Pairs merged during Stage B\textquotesingle\ duplicate detection fall below $\theta_{B'} = 0.01$. Remaining groups are ranked by mean absolute deviation from peer consensus in Stage C\textquotesingle. \textbf{(b)} Per-point uncertainty across consensus tiers (\S\ref{sec:tiers}). Dashed line: $\sigma$-floor $= 0.012$; dotted lines: per-tier medians. Gold is concentrated between the floor and ${\sim}0.03$ log S; Bronze tails past $\sigma > 0.2$. Coverage is near-identical (${\sim}77\,\%$) --- the difference is about \emph{quality} of uncertainty.}
\label{fig:curation_and_tiers}
\end{figure}

In order to maximize the data integrity and reliability we have performed a multi-step curation pipeline applied to raw \textsc{BigSolDB}~v2.1 dataset. This included:

\begin{itemize} [leftmargin=1.2em,itemsep=2pt,topsep=2pt]
  \item \textbf{Raw data audit.} This included reconstruction of \logS{} for the entries with missing values, canonicalization of SMILES strings and removing the exact duplicate rows (identical canonical solute, canonical solvent, $T$ rounded to $0.1$\,K, \logS{} values).
  \item \textbf{Source integrity analysis.} During this section groups with the same [canonical solute;canonical solvent] pairs were further analyzed.  As some of the duplicate measurements can not be spotted upon the exact duplicate exclusion, thus deflating inter-lab statistics, we have performed additional analysis on those that are not bit-exact but share very similar values. The opposite cases were the groups for which excessively big inter-lab variations in a same solubility measurements occur. These were resolved manually by either correcting the mistakes of the \textsc{BigSolDB}~v2.1 extraction or flagging the data as unreliable (stated as bad DOIs hereinafter). The overall inter-lab discrepancy stats are presented in Figures \ref{fig:stages_bc} and \ref{fig:tier_sigma}. 
  \item \textbf{Cleaning waterfall.} The cleaning waterfall included removing the bad DOIs, invalid / polymer solvent SMILES, salts/mixtures/excessively big ($M_w > 1000$\,Da) solutes, recovering absent \logS{} values, removing extremes ($\log_{10} S \notin [-15, 2]$) and near-duplicate exclusion. All these procedures were aimed at either correcting the definite mistakes or removing obvious outlier points, increasing the overall data quality and integrity.
\end{itemize}
 The detailed procedure for each step of data curation is presented in \S\ref{app:data_curation}. After all the procedures, the final \scthree{} dataset subjected to further benchmark construction contains {101\,535} measurements over
{1\,327} solutes, {206} solvents, and {1\,493} DOIs
($T \in [243, 426]$\,K; $\log_{10} S \in [-7.56, +1.99]$), of which
{746} (solute,\,solvent) pairs have $\geq 2$ independent sources and
form the multi-source pool for \S\ref{sec:aleatoric}.

\subsection{Aleatoric limit}
\label{sec:aleatoric}

The aleatoric limit---the disagreement between independent measurements of a (solute, solvent, $T$) point---forms an irreducible error floor \citep{attia2025data}. While the widely cited 0.6--0.8 log S figure is often assumed to be this ceiling, we show that it actually reflects a heavy-tailed $P_{90}$--$P_{95}$ worst-case; expected inter-lab disagreement is roughly $6\times$ smaller.

\paragraph{Mathematical framework.}
For a (solute,\,solvent) pair $p$ with independence groups $\mathcal{G}_p$, the pair MAE averages
the interpolated disagreement $|f_i(T) - f_j(T)|$ between fitted curves for all group pairs
$i < j$ over a 1\,K reference grid $\mathcal{T}$. The global aleatoric limit
$\varepsilon_{\mathrm{aleatoric}}$ is the mean pair MAE across the multi-source pool $\mathcal{P}$:
\begin{equation}\label{eq:aleatoric}
    \varepsilon_{\mathrm{aleatoric}} =
    \frac{1}{|\mathcal{P}|}\sum_{p \in \mathcal{P}}
    \left[
        \frac{1}{\binom{|\mathcal{G}_p|}{2}}
        \sum_{i < j \in \mathcal{G}_p}
        \mathrm{mean}_{T \in \mathcal{T}}\,\bigl|f_i(T) - f_j(T)\bigr|
    \right].
\end{equation}

\textbf{Primary vs.\ inclusive.}
We exclude Low-Reliability DOIs (\S\ref{app:source_integrity}) from tier consensus labels (\S\ref{sec:tiers}). We report two aleatoric limits: \emph{primary} (excluding these DOIs, matching the benchmark's tier pool) and \emph{inclusive} (retaining them for literature comparison).
\begin{table}[h]
\centering\footnotesize
\caption{Primary and inclusive $\varepsilon_{\mathrm{aleatoric}}$. 95\,\% confidence intervals
are from 5\,000 bootstrap resamples over the multi-source pair pool.}
\label{tab:aleatoric}
\resizebox{\linewidth}{!}{%
\begin{tabular}{@{}l r c c c c c c@{}}
\toprule
& {$n$ pairs} & {mean $= \varepsilon_{\mathrm{aleatoric}}$} & {95\,\% CI} & {median} & {$P_{90}$} & {$P_{95}$} & {RMSE} \\
\midrule
\textbf{Primary} (LR-DOIs excluded) & 481 & 0.106 & {[0.093, 0.120]} & 0.046 & 0.258 & 0.385 & 0.182 \\
Inclusive                           & 511 & 0.158 & {[0.132, 0.186]} & 0.050 & 0.385 & 0.627 & 0.356 \\
\bottomrule
\end{tabular}%
}
\end{table}

\textbf{Per-solvent heterogeneity.}
$\varepsilon_{\mathrm{aleatoric}}$ varies substantially across solvents
(Figure~\ref{fig:per_solvent_aleatoric}, Appendix~\ref{app:data_curation}):
DMF is tightest at $0.029$~log~S (10 pairs), while water---despite its large
data volume---is highest among common solvents at $0.110$~log~S (65 pairs),
driven by a thicker tail rather than a higher median ($0.061$). A single global
$\varepsilon_{\mathrm{aleatoric}}$ therefore under-characterises difficult
solvents and over-characterises well-behaved ones, motivating PS-RMSE
(\S\ref{sec:metrics}). The commonly cited $0.6$--$0.8$~log~S figure of
\citet{palmer2014dataquality} coincides with our $P_{90}$--$P_{95}$
($0.39$--$0.63$) and RMSE ($0.36$), confirming it reflects worst-case noise
rather than the expected floor ($\varepsilon_{\mathrm{aleatoric}} \approx 0.11$).

%% file: sections/03.benchmark_design.tex
\section{Benchmark design}
\label{sec:benchmark}


\scthree{} aims to answer three distinct generalisation questions: can a model predict (i) a new $(\mathrm{solute}, \mathrm{solvent})$ pair in a familiar solvent,(ii) the same solute in a solvent it has never seen, and (iii) a new solute against consensus-calibrated labels with per-point uncertainty $\sigma$? Thus \scthree{} delivers three things in concert: tiered evaluation sets with consensus labels and per-point uncertainty~$\sigma$ (\S\ref{sec:tiers}), leakage-checked splits across three distinct generalisation axes (\S\ref{sec:splits}), and a metric suite suited to the multi-solvent structure of the data (\S\ref{sec:metrics}). 
The tier-test pool is drawn from the {481} multi-source pairs of the cleaned dataset; its complement forms the training pool, further
divided by solvent frequency into in-distribution (Train, Eval) and out-of-distribution (OOD) sets.  Aggregate RMSE is insufficient to score this structure - we characterise why empirically and define the five-metric suite in \S\ref{sec:metrics}.

\subsection{Tier construction}
\label{sec:tiers}

The tiers provide calibrated ground truth with per-point $\sigma$, built
from the {481} multi-source pairs. For each pair we compute the
pair MAE ($\binom{|\mathcal{G}_p|}{2}^{-1} \sum_{i < j \in
\mathcal{G}_p} \mathrm{mean}_{T \in \mathcal{T}}\,\bigl|f_i(T) -
f_j(T)\bigr|$) and assign it to every tier whose threshold it satisfies:
Gold ($\leq 0.1$~log~S, ${\sim}\epsA$), Silver ($\leq 0.2$~log~S,
${\sim}2\epsA$), and Bronze ($\leq 0.5$~log~S, ${\sim}5\epsA$). The
tiers are nested: $\text{Gold} \subset \text{Silver} \subset
\text{Bronze}$; a tighter tier means better-characterised ground truth,
not a harder modelling task.

\textbf{Consensus labels.}
For each $(\mathrm{solute}, \mathrm{solvent}, T)$ triple in a tier pair,
we evaluate every contributing independence group's Apelblat /
van't-Hoff fit at $T$ and define the consensus label as the mean,
excluding Low-Reliability DOI groups. The per-point uncertainty $\sigma$
is the standard deviation across the same evaluations; when fewer than
two groups contribute, $\sigma$ is undefined and the row carries a hard
label ({14.4}\,\% of tier rows).

\textbf{Tier statistics.}
Table~\ref{tab:tiers} reports size and $\sigma$ distribution per tier.
Median $\sigma$ ranges from {0.019} (Gold) to {0.031}
(Bronze); Gold's tail is much shorter ($P_{95} = {0.078}$ vs.\
Bronze $P_{95} = {0.245}$). Figure~\ref{fig:tier_sigma} shows
the full distribution.

\begin{table}[h]
\centering\small
\caption{SC$^{3}$ tier composition and $\sigma$ statistics. Nested:
Gold $\subset$ Silver $\subset$ Bronze. $\sigma$-coverage is the
fraction of tier rows with defined $\sigma$;
$\varepsilon$-RMSE\,$=\sqrt{\langle\sigma^{2}\rangle}$ is the
aleatoric floor on that subset.}
\label{tab:tiers}
\begin{tabular}{@{}lrrrr ccccc@{}}
\toprule
Tier   & rows  & pairs & solutes & solvents
       & $\sigma$-cov. & median $\sigma$ & mean $\sigma$ & $P_{95}\ \sigma$ & $\varepsilon$-RMSE \\
\midrule
Gold   & 4\,507 & 335 & 129 & 26 & 76.5\%  & 0.019 & 0.028 & 0.078 & 0.037 \\
Silver & 5\,475 & 400 & 141 & 27 & 76.9\%  & 0.024 & 0.042 & 0.130 & 0.059 \\
Bronze & 6\,331 & 469 & 148 & 30 & 77.1\%  & 0.031 & 0.066 & 0.245 & 0.102 \\
\bottomrule
\end{tabular}
\end{table}

\subsection{Training, evaluation, and OOD Splits}
\label{sec:splits}

The {148} solutes appearing in any tier are fully excluded from the training pool at the $(\mathrm{solute}, \mathrm{solvent}, T)$
level; the remaining {80\,312} rows are split by solvent frequency.  The top-25 solvents by row count (covering {85.1\,\%} of the training pool) form the in-distribution region: {10\,\%} of each solvent's solute list is held out as \textbf{Eval}, the remainder is \textbf{Train}.  The remaining {181} solvents form \textbf{OOD}, on which any model is
extrapolating across the solvent axis by construction. Table~\ref{tab:splits} summarises the composition. 

\begin{table}[t]
\centering\small
\caption{SC$^{3}$ splits and tiers.  All six evaluation subsets have
zero solute overlap with training (see Table~\ref{tab:antileakage}).
Solvent overlap with Train is intentional for Eval (in-distribution)
and tiers (calibrated labels); OOD has zero solvent overlap with Train
by construction.}
\label{tab:splits}
\begin{tabular}{@{}llrrrr@{}}
\toprule
& Role                           & rows     & solutes & solvents & (solute, solvent) pairs \\
\midrule
\multicolumn{6}{l}{\textit{Training pool ({80\,312} rows after tier exclusion)}} \\
Train  & training set                   & 61\,403  & 1\,144  & 25       & 6\,840 \\
Eval   & new-pair, in-distribution      &  6\,969  &    534  & 25       &    771 \\
OOD    & new-solvent (151 unseen)       & 11\,940  &    586  & 161      &  1\,450 \\
\midrule
\multicolumn{6}{l}{\textit{Tier test pool (solutes fully absent from training pool)}} \\
Gold   & tight consensus, $\sigma$-cal.  &  4\,507  &    129  &  26      &    335 \\
Silver & looser consensus, $\sigma$-cal. &  5\,475  &    141  &  27      &    400 \\
Bronze & broadest, $\sigma$-cal.         &  6\,331  &    148  &  30      &    469 \\
\bottomrule
\end{tabular}
\end{table}
The three evaluation sets probe distinct generalisation axes:
\textbf{Eval} tests new $(\mathrm{solute}, \mathrm{solvent})$ pairs
in already-seen solvents (solute interpolation);
\textbf{OOD} tests the {161} long-tail solvents, none of which
appear in training (solvent extrapolation, reported with per-solvent
split-out in \S\ref{sec:metrics});
and \textbf{Gold / Silver / Bronze} test new solutes against consensus labels with calibrated $\sigma$ (solute-level out-of-distribution, since all {148} tier solutes are absent from Train, Eval, and OOD). Every pair-level overlap that must be zero is zero
(Table~\ref{tab:antileakage}); the {520}-solute overlap between Train and Eval is expected, as Eval is a held-out slice of the in-distribution solute--solvent grid. All required pair-level overlaps between splits are zero; the sole expected exception is the {520}-solute overlap between Train
and Eval, which shares solutes but no (solute, solvent) pairs by construction (full verification in Table~\ref{tab:antileakage}, Appendix~\ref{app:data_curation}).



\subsection{Metrics}
\label{sec:metrics}

With the splits in place (§\ref{sec:benchmark}), we turn to scoring. Reporting aggregate RMSE works for single-solvent regression but fails
here for four distinct reasons rooted in the multi-solvent structure
of the data.  We characterise those reasons empirically on \scthree{},
then define the five-metric suite we recommend reporting for every
model.\\
\textbf{Per-solvent location shift.}
Across the {206} cleaned solvents, per-solvent log~S means span
\textbf{{7.98} log units} (minimum {-6.93}, maximum
{+1.05}); nine orders of magnitude in equivalent
concentration.  Figure~\ref{fig:solvent_dist} shows the per-solvent
log~S distribution for the top-20 solvents by row count, ordered by
their per-solvent mean.  The densities overlap substantially in the
middle of the range, but their locations differ by up to 7 log units. \\
\textbf{Variance decomposition.}
A one-way ANOVA of cleaned log~S on solvent identity attributes
\textbf{{11.7\,\%}} of total variance to the between-solvent
location shift ($F = {69.9}$ across {206} solvents,
$p < 10^{-300}$).  on OOD where long-tail solvents include extremes
(e.g.\ 1,1-dichloroethane, mean log\,S $= -6.93$) the between-solvent fraction rises to \textbf{{22.1\,\%}}
Within each tier the fraction is closer to {9.3}--{9.9\,\%}.\\
\textbf{Dummy-baseline R$^2$.}
We quantify how much of this variance is ``free'' for a trivial model.
A predictor that returns the \emph{per-solvent training mean} (no
solute information, no temperature) scores $R^{2} = {+}0.053$ on Eval but $R^{2} = {-}0.062$ on OOD, where it falls back to the grand mean across {161} unseen solvents; on tiers $R^2 \in [-0.003, +0.017]$.
Aggregate metrics that do not strip between-solvent variance reward
memorising solvent means rather than genuine solute prediction. --- the grand mean
is actively worse than no prediction at all on the long-tail solvents.\\
\textbf{Count domination.}
The top-5 solvents (ethanol, methanol, 2-propanol, ethyl acetate,
1-propanol) account for {37.5}\,\% of cleaned rows and
{44.2}\,\% of Train; the top-25 cover 100\,\% of Train/Eval
by construction.  Aggregate RMSE is therefore functionally a metric
on five solvents, hiding performance on the OOD set.\\
\textbf{MAPE exhibits numerical instability.}
{5.6}\,\% of rows have $|\log_{10} S| < 0.1$ and
{28.5}\,\% have $|\log_{10} S| < 0.5$, which may make MAPE unstable. It is thus excluded from the headline suite.\\
\textbf{Heavy-tailed labels.}
The $|y - \bar{y}|$ distribution has mean/median ratio
${1.19}$ ($P_{95} = {2.27}$, $P_{99} =
{3.43}$).  RMSE is tail-sensitive in this heavy regime,
supporting median absolute error as a robust complement.

\paragraph{The SC$^3$ metric suite}
Aggregate metrics like RMSE, MAE, and MedAE suffer from severe count bias in the multi-solvent setting, as they are dominated by high-frequency solvents (\S\ref{sec:metrics}). We therefore adopt \textbf{PS-RMSE} as our headline metric to equalize solvent contributions, and \textbf{Z-RMSE} to measure performance directly against the aleatoric noise limit. Table~\ref{tab:metrics_and_domain} defines the metric suite; notably, Z-RMSE is computed only on the ${\sim}77\,\%$ of tier rows where per-point $\sigma$ is defined.

\begin{table}[h]
\centering
\scriptsize
\renewcommand{\arraystretch}{1.2}
\caption{The SC$^{3}$ metric suite. Standard aggregate metrics suffer from count bias; PS-RMSE cancels between-solvent shifts, and Z-RMSE normalizes by calibrated uncertainty ($\sigma_i$).}
\label{tab:metrics_and_domain}
\begin{tabular}{@{} l l p{7.5cm} @{}}
\toprule
\textbf{Metric} & \textbf{Definition} & \textbf{Detail} \\
\midrule
RMSE & $\sqrt{ \tfrac{1}{n} \sum_i (\hat y_i-y_i)^2 }$ & Standard baseline; count-weighted and dominated by high-frequency solvents. \\
MAE & $\tfrac{1}{n} \sum_i |\hat y_i-y_i|$ & Less tail-sensitive (outliers do not contribute quadratically); holds count bias. \\
MedAE & $\mathrm{median}_i \, |\hat y_i-y_i|$ & Robust summary under heavy-tailed inter-lab disagreement; holds count bias. \\
\textbf{PS-RMSE} & $\tfrac{1}{|\mathcal{S}|} \sum_{s} \mathrm{RMSE}_s$ & \textbf{Headline metric.} Cancels between-solvent location shifts; every solvent contributes equally. Converges to true within-solvent prediction error. \\
\textbf{Z-RMSE} & $\sqrt{ \tfrac{1}{n_\sigma} \sum\limits_{i:\sigma_i\,\mathrm{def.}} \!\left( \tfrac{\hat y_i-y_i}{\sigma_i} \right)^{\!2} }$ & Normalizes prediction error by aleatoric uncertainty. A value of ${\sim}1$ means the model operates at the measurement-noise limit. \\
\textit{MAPE} & $\tfrac{1}{n} \sum_i \tfrac{|\hat y_i-y_i|}{|y_i|}$ & Diagnostic only. Denominator is unstable because $5.6\,\%$of dataset rows have $|\!\log_{10} S| < 0.1$. \\
\bottomrule
\end{tabular}
\end{table}

%% file: sections/04.main_results.tex

\section{Baselines and main results}
\label{sec:baselines}

Solubility prediction has cycled through three modelling generations;
$\mathrm{SC}^{3}$ evaluates a representative of each under identical
conditions. The first is thermodynamic and additive: the Yalkowsky GSE
\citep{yalkowsky1980solubility}, UNIFAC group-contribution models
\citep{fredenslund1975unifac}, Abraham LFER \citep{abraham1993scales}, and
Delaney's ESOL \citep{delaney2004esol}. These define the chemical
axes---polarity, surface area, H-bond counts, melting point---that any
data-driven model must recover. The second generation replaced linear
regression with gradient-boosted trees and random forests
\citep{breiman2001randomforests,chen2016xgboost,ke2017lightgbm,prokhorenkova2018catboost};
RDKit or Mordred descriptors with a tuned ensemble remain highly competitive
\citep{boobier2020machine,tayyebi2023prediction,ramani2026dissolvr}. A parallel
branch substituted circular fingerprints (Morgan/ECFP4
\citep{rogers2010extended}) with the same ensembles or Tanimoto-kernel methods
\citep{ralaivola2005graph}.

The third generation introduces learned representations. Descriptor-based deep
regressors---FastProp \citep{burns2025fastprop} (feedforward on $\sim$1500
Mordred features), its solubility-specialised variant
FastSolv \citep{attia2025data}, and
Dissolvr \citep{ramani2026dissolvr} (RDKit + MOSE/Joback/Abraham
blocks)---match or narrowly beat trees on in-distribution splits at a fraction
of training cost. Graph neural networks form the dominant deep paradigm:
GCN \citep{kipf2017gcn}, GAT \citep{velickovic2018gat}, and
GIN \citep{xu2019gin} supply backbones; Chemprop's directed message-passing
network \citep{yang2019chemprop,heid2024chemprop} is the de-facto solubility
benchmark; and explicit solute--solvent encoders---CIGIN \citep{pathak2021cigin},
MolMerger \citep{ramani2024molmerger}, the SE(3)-equivariant
Solvaformer \citep{broadbent2025solvaformer}, and
SolubNet \citep{chen2023solubnet} (topology-adaptive convolutions with
layer-wise relevance readout)---encode the inductive bias that solubility is a
property of a \emph{pair}. Foundation-scale baselines complete the catalogue:
SolTranNet \citep{francoeur2021soltrannet} (3393-parameter molecule-attention
transformer), Uni-Mol2 \citep{ji2024unimol2} (1.1B-parameter 3D-pretrained,
frozen and fine-tuned), and ChemFM \citep{cai2026chemfm} (3B-parameter SMILES
language model) probe whether generic pretraining transfers to multi-solvent
solubility.

Thirty-one models drawn from these six families are evaluated under one
protocol.  Every method consumes the same $(\text{solute}, \text{solvent}, T)$
input, sees identical splits (\S\ref{sec:splits}), and is scored on the metric
suite of \S\ref{sec:metrics}.  Hyperparameters are selected once on
\textsc{Eval} via a fixed grid for tree ensembles and a modest manual sweep
for the deep models---defaults for the foundation models---and frozen across all subsequent runs; reported numbers are mean$\,\pm\,$std over five seeds
$\{42, 101, 123, 456, 789\}$.

\input{tables/main_results}      

Four patterns dominate Table~\ref{tab:main_results}. First, descriptors
decisively beat fingerprints: CatBoost on RDKit reaches $0.593$ Bronze
PS-RMSE versus $0.766$ on Morgan ECFP4, a $0.17\,\log S$ gap recurring
across all four tree ensembles---traced in \S\ref{sec:interpretability} to
fingerprints lacking a representation that distinguishes water from a
long-chain alcohol. Second, deep descriptor models (FastProp, FastSolv,
MLP-on-RDKit) match or narrowly beat trees on \textsc{Eval} (FastProp
$0.465$ vs.\ CatBoost $0.477$) but trail by $0.06$--$0.11$ on OOD and
$0.13$--$0.15$ on Bronze---a generalisation gap, not a fitting one. Third,
plain dual-encoder GNNs (GCN, GAT, GIN) sit ${\sim}0.4\,\log S$ above
descriptor methods on Bronze; solvent-aware variants (Chemprop, MolMerger,
Solvaformer, SolubNet, RiLOOD) close most of that gap to $0.13$--$0.21$
above the best tree, indicating the solute--solvent encoder matters more
than the message-passing backbone. Fourth, foundation-model fine-tunes
(Uni-Mol2, SolTranNet, ChemFM) land at $0.71$--$0.76$ Bronze and additive
physics baselines (Abraham LFER, ESOL, GSE) at $0.80$--$0.83$. The
headline: RDKit features with a tuned tree ensemble remain the
configuration to beat at ${\approx}\,5\times
\varepsilon_{\mathrm{aleatoric}}$ on Bronze PS-RMSE; no deep alternative
closes this gap---including ChemFM, whose $0.437$ \textsc{Eval} RMSE is
the table's lowest but whose $0.714$ Bronze still trails by
$0.15\,\log S$.

%% file: tables/main_results.tex

\definecolor{first}{HTML}{C6EFCE}
\definecolor{second}{HTML}{DAEFC6}
\definecolor{third}{HTML}{EEF5DD}

\providecommand{\fst}[1]{\cellcolor{first}\textbf{#1}}
\providecommand{\snd}[1]{\cellcolor{second}#1}
\providecommand{\trd}[1]{\cellcolor{third}#1}
\providecommand{\old}[1]{\textcolor{gray!75}{#1}}

\begin{table*}[t]
  \caption{\scthree{} benchmark.  Mean{\scriptsize$\pm$std} over 5 seeds;
  PS-RMSE is the primary metric (top 3 per column among all rows
  highlighted, best in bold green).}
  \label{tab:main_results}
  \centering
  \scriptsize
  \setlength{\tabcolsep}{2.5pt}
  \renewcommand{\arraystretch}{0.92}
  \begin{tabular}{@{}l l cc c ccc c@{}}
    \toprule
    & & \multicolumn{2}{c}{Eval} & OOD & \multicolumn{3}{c}{PS-RMSE\,$\downarrow$} & Z\,$\downarrow$ \\
    \cmidrule(lr){3-4} \cmidrule(lr){6-8}
    Method & Repr. & RMSE & $R^2$ & RMSE & Gold & Silver & Bronze & Bronze \\
    \midrule
    \multicolumn{9}{@{}l}{\textit{Descriptor + tree ensembles}}\\
    LightGBM      & RDKit    & 0.493\,$\pm$\,0.003 & 0.816\,$\pm$\,0.003 & \snd{0.611\,$\pm$\,0.004} & \fst{0.604\,$\pm$\,0.008} & \fst{0.561\,$\pm$\,0.009} & \fst{0.561\,$\pm$\,0.010} & \fst{35.28\,$\pm$\,0.31} \\
    Abraham\,ML   & Abraham  & 0.548\,$\pm$\,0.004 & 0.772\,$\pm$\,0.004 & 0.686\,$\pm$\,0.005 & 0.879\,$\pm$\,0.016 & 0.814\,$\pm$\,0.014 & 0.835\,$\pm$\,0.017 & 44.31\,$\pm$\,0.30 \\
    CatBoost      & RDKit    & 0.477\,$\pm$\,0.008 & 0.827\,$\pm$\,0.006 & \fst{0.611\,$\pm$\,0.005} & \trd{0.635\,$\pm$\,0.029} & \trd{0.597\,$\pm$\,0.028} & \trd{0.593\,$\pm$\,0.032} & \trd{36.07\,$\pm$\,0.53} \\
    Dissolvr      & Domain   & 0.490\,$\pm$\,0.004 & 0.818\,$\pm$\,0.003 & \trd{0.619\,$\pm$\,0.003} & \snd{0.623\,$\pm$\,0.009} & \snd{0.578\,$\pm$\,0.004} & \snd{0.576\,$\pm$\,0.005} & \snd{36.05\,$\pm$\,0.50} \\
    XGBoost       & RDKit    & 0.494\,$\pm$\,0.010 & 0.815\,$\pm$\,0.007 & 0.641\,$\pm$\,0.013 & 0.651\,$\pm$\,0.015 & 0.614\,$\pm$\,0.014 & 0.604\,$\pm$\,0.016 & 36.60\,$\pm$\,0.98 \\
    Random Forest & RDKit    & 0.517\,$\pm$\,0.001 & 0.797\,$\pm$\,0.001 & 0.624\,$\pm$\,0.001 & 0.683\,$\pm$\,0.004 & 0.629\,$\pm$\,0.002 & 0.634\,$\pm$\,0.002 & 36.72\,$\pm$\,0.12 \\
    Tayyebi       & Mordred  & 0.573\,$\pm$\,0.001 & 0.752\,$\pm$\,0.001 & 0.717\,$\pm$\,0.001 & 0.795\,$\pm$\,0.009 & 0.728\,$\pm$\,0.006 & 0.738\,$\pm$\,0.005 & 40.34\,$\pm$\,0.23 \\
    \midrule
    \multicolumn{9}{@{}l}{\textit{Fingerprint-based}}\\
    CatBoost      & Morgan   & 0.546\,$\pm$\,0.002 & 0.774\,$\pm$\,0.002 & 0.728\,$\pm$\,0.008 & 0.878\,$\pm$\,0.018 & 0.819\,$\pm$\,0.017 & 0.766\,$\pm$\,0.021 & 42.78\,$\pm$\,0.60 \\
    XGBoost       & Morgan   & 0.565\,$\pm$\,0.004 & 0.758\,$\pm$\,0.003 & 0.778\,$\pm$\,0.002 & 0.985\,$\pm$\,0.018 & 0.918\,$\pm$\,0.012 & 0.829\,$\pm$\,0.014 & 47.37\,$\pm$\,0.58 \\
    Random Forest & Morgan   & 0.578\,$\pm$\,0.001 & 0.747\,$\pm$\,0.001 & 0.756\,$\pm$\,0.001 & 0.817\,$\pm$\,0.003 & 0.782\,$\pm$\,0.004 & 0.719\,$\pm$\,0.004 & 44.72\,$\pm$\,0.03 \\
    LightGBM      & Morgan   & 0.532\,$\pm$\,0.004 & 0.785\,$\pm$\,0.003 & 0.778\,$\pm$\,0.006 & 0.924\,$\pm$\,0.013 & 0.862\,$\pm$\,0.010 & 0.792\,$\pm$\,0.011 & 45.27\,$\pm$\,0.38 \\
    GP            & Tanimoto & 0.627\,$\pm$\,0.013 & 0.702\,$\pm$\,0.012 & 0.958\,$\pm$\,0.012 & 1.038\,$\pm$\,0.054 & 0.967\,$\pm$\,0.040 & 0.909\,$\pm$\,0.028 & 50.17\,$\pm$\,1.52 \\
    \midrule
    \multicolumn{9}{@{}l}{\textit{Deep learning (descriptors)}}\\
    FastProp      & RDKit    & \fst{0.465\,$\pm$\,0.006} & \fst{0.837\,$\pm$\,0.004} & 0.675\,$\pm$\,0.013 & 0.768\,$\pm$\,0.041 & 0.732\,$\pm$\,0.034 & 0.689\,$\pm$\,0.031 & 38.61\,$\pm$\,1.47 \\
    FastSolv      & RDKit    & \trd{0.471\,$\pm$\,0.006} & \trd{0.832\,$\pm$\,0.004} & 0.708\,$\pm$\,0.008 & 0.791\,$\pm$\,0.063 & 0.748\,$\pm$\,0.045 & 0.693\,$\pm$\,0.039 & 38.28\,$\pm$\,0.93 \\
    MLP           & RDKit    & \snd{0.470\,$\pm$\,0.004} & \snd{0.833\,$\pm$\,0.003} & 0.720\,$\pm$\,0.007 & 0.813\,$\pm$\,0.030 & 0.768\,$\pm$\,0.034 & 0.708\,$\pm$\,0.036 & 40.00\,$\pm$\,1.09 \\
    \midrule
    \multicolumn{9}{@{}l}{\textit{Graph neural networks}}\\
    Chemprop      & D-MPNN   & \old{0.471\,$\pm$\,0.010} & \old{0.852\,$\pm$\,0.007} & \old{0.680\,$\pm$\,0.014} & \old{0.802\,$\pm$\,0.047} & \old{0.747\,$\pm$\,0.035} & \old{0.688\,$\pm$\,0.035} & \old{47.33\,$\pm$\,2.31} \\
    Solvaformer   & SE(3)    & \old{0.510\,$\pm$\,0.015} & \old{0.827\,$\pm$\,0.010} & \old{0.768\,$\pm$\,0.027} & \old{0.877\,$\pm$\,0.063} & \old{0.788\,$\pm$\,0.061} & \old{0.722\,$\pm$\,0.033} & \old{45.12\,$\pm$\,3.04} \\
    GCN           & Dual enc & 0.599\,$\pm$\,0.016 & 0.728\,$\pm$\,0.015 & 0.776\,$\pm$\,0.031 & 1.066\,$\pm$\,0.039 & 1.014\,$\pm$\,0.049 & 0.979\,$\pm$\,0.042 & 55.15\,$\pm$\,2.88 \\
    SolubNet      & TAGConv  & \old{0.524\,$\pm$\,0.024} & \old{0.817\,$\pm$\,0.017} & \old{0.741\,$\pm$\,0.022} & \old{0.942\,$\pm$\,0.049} & \old{0.839\,$\pm$\,0.045} & \old{0.768\,$\pm$\,0.029} & \old{50.62\,$\pm$\,2.18} \\
    RiLOOD        & CIGIN    & \old{0.557\,$\pm$\,0.036} & \old{0.793\,$\pm$\,0.027} & \old{0.969\,$\pm$\,0.016} & \old{0.965\,$\pm$\,0.127} & \old{0.869\,$\pm$\,0.089} & \old{0.769\,$\pm$\,0.085} & \old{50.52\,$\pm$\,3.21} \\
    GAT           & Dual enc & 0.567\,$\pm$\,0.028 & 0.756\,$\pm$\,0.024 & 0.787\,$\pm$\,0.028 & 1.082\,$\pm$\,0.054 & 1.012\,$\pm$\,0.061 & 0.963\,$\pm$\,0.054 & 54.77\,$\pm$\,2.40 \\
    GIN           & Dual enc & 0.553\,$\pm$\,0.011 & 0.769\,$\pm$\,0.009 & 0.784\,$\pm$\,0.028 & 1.020\,$\pm$\,0.024 & 0.966\,$\pm$\,0.034 & 0.935\,$\pm$\,0.032 & 55.27\,$\pm$\,3.52 \\
    MolMerger     & Merged   & \old{0.491\,$\pm$\,0.010} & \old{0.840\,$\pm$\,0.006} & \old{0.692\,$\pm$\,0.022} & \old{0.853\,$\pm$\,0.058} & \old{0.794\,$\pm$\,0.040} & \old{0.707\,$\pm$\,0.051} & \old{41.42\,$\pm$\,1.74} \\
    \midrule
    \multicolumn{9}{@{}l}{\textit{Foundation / pretrained}}\\
    Uni-Mol2 + CB & 3D + Tree & 0.494\,$\pm$\,0.004 & 0.815\,$\pm$\,0.003 & 0.732\,$\pm$\,0.007 & 0.817\,$\pm$\,0.030 & 0.773\,$\pm$\,0.023 & 0.756\,$\pm$\,0.025 & 38.71\,$\pm$\,0.89 \\
    Uni-Mol2      & 3D repr   & 0.497\,$\pm$\,0.004 & 0.813\,$\pm$\,0.003 & 0.659\,$\pm$\,0.011 & 0.849\,$\pm$\,0.024 & 0.807\,$\pm$\,0.021 & 0.744\,$\pm$\,0.020 & 40.49\,$\pm$\,0.69 \\
    SolTranNet    & Char-TF   & 0.507\,$\pm$\,0.010 & 0.805\,$\pm$\,0.008 & 0.742\,$\pm$\,0.019 & 0.823\,$\pm$\,0.050 & 0.782\,$\pm$\,0.034 & 0.735\,$\pm$\,0.033 & 43.49\,$\pm$\,1.38 \\
    ChemFM        & LoRA      & \old{0.437\,$\pm$\,0.012} & \old{0.873\,$\pm$\,0.009} & \old{0.708\,$\pm$\,0.024} & \old{0.785\,$\pm$\,0.045} & \old{0.741\,$\pm$\,0.032} & \old{0.714\,$\pm$\,0.028} & \old{40.12\,$\pm$\,1.55} \\
    \midrule
    \multicolumn{9}{@{}l}{\textit{Physics / analytical}}\\
    UNIFAC        & Grp-contr & \old{1.417\,$\pm$\,0.001} & \old{$-$0.335\,$\pm$\,0.001} & \old{1.419\,$\pm$\,0.002} & \old{1.087\,$\pm$\,0.005} & \old{1.195\,$\pm$\,0.005} & \old{1.051\,$\pm$\,0.003} & \old{65.96\,$\pm$\,0.12} \\
    Abraham LFER  & 5-param   & 0.965\,$\pm$\,0.000 & 0.295\,$\pm$\,0.000 & 1.023\,$\pm$\,0.000 & 0.852\,$\pm$\,0.000 & 0.810\,$\pm$\,0.000 & 0.804\,$\pm$\,0.000 & 41.95\,$\pm$\,0.00 \\
    ESOL          & Linear    & 1.094\,$\pm$\,0.000 & 0.094\,$\pm$\,0.000 & 1.083\,$\pm$\,0.000 & 0.825\,$\pm$\,0.000 & 0.829\,$\pm$\,0.000 & 0.814\,$\pm$\,0.000 & 45.49\,$\pm$\,0.00 \\
    GSE           & $-\log P$ & 1.112\,$\pm$\,0.000 & 0.064\,$\pm$\,0.000 & 1.094\,$\pm$\,0.000 & 0.828\,$\pm$\,0.000 & 0.846\,$\pm$\,0.000 & 0.829\,$\pm$\,0.000 & 49.12\,$\pm$\,0.00 \\
    \bottomrule
  \end{tabular}
\end{table*}

%% file: sections/05.ablations.tex
\section{Analysis \& future directions}
\label{sec:ablations}

\input{sections/05a.interpretations}

\begin{figure*}[t]
\centering
\includegraphics[width=\linewidth]{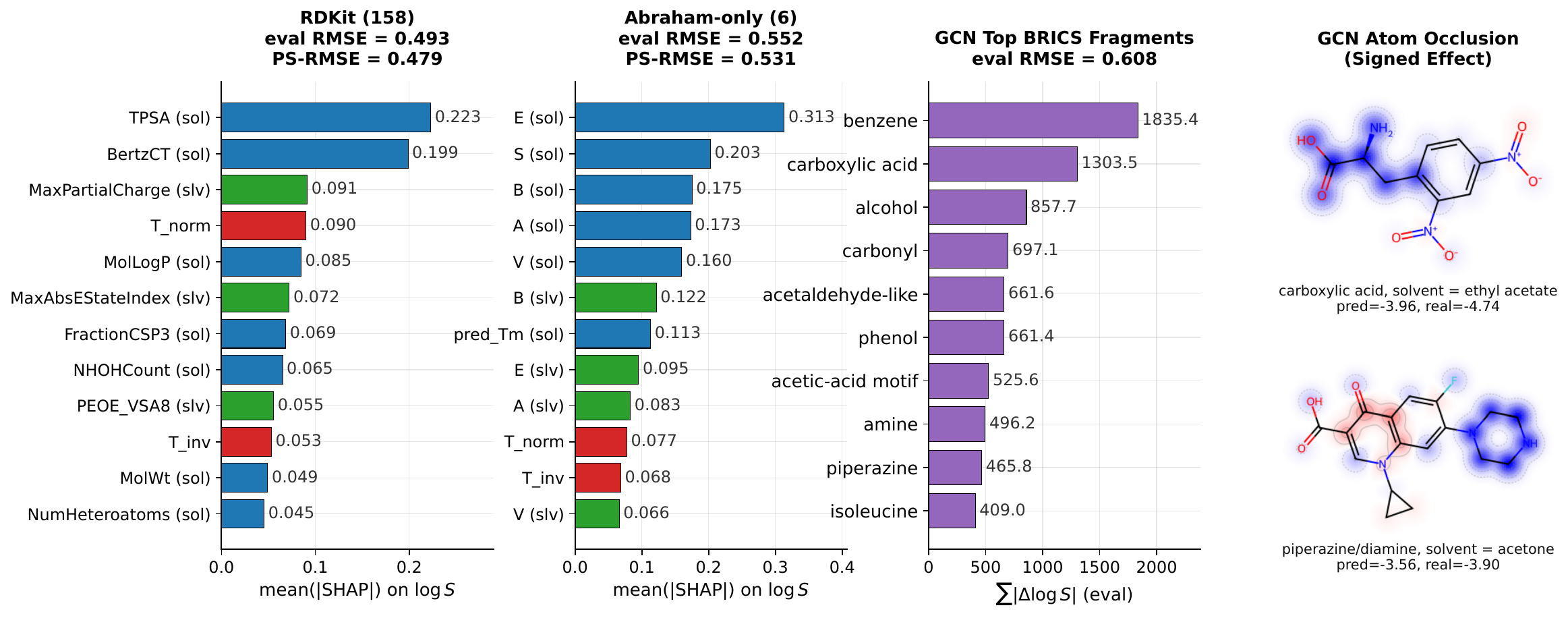}
\caption{\textbf{Interpretability ablation.} (Left to right) Top-12 RDKit SHAP features, Top-12 Abraham-only SHAP features, GCN top BRICS fragment occlusions, and GCN Signed Atom Occlusion Halos for selected molecules (where red raises and blue lowers the predicted solubility; molecules shown top to bottom: carboxylic acid derivative, piperazine derivative). SHAP feature blocks are coloured by solute (blue), solvent (green), and temperature (red).}
\label{fig:row1_interp}
\end{figure*}

\input{sections/05b.data_scaling}

\begin{figure*}[t]
\centering
\includegraphics[width=\linewidth]{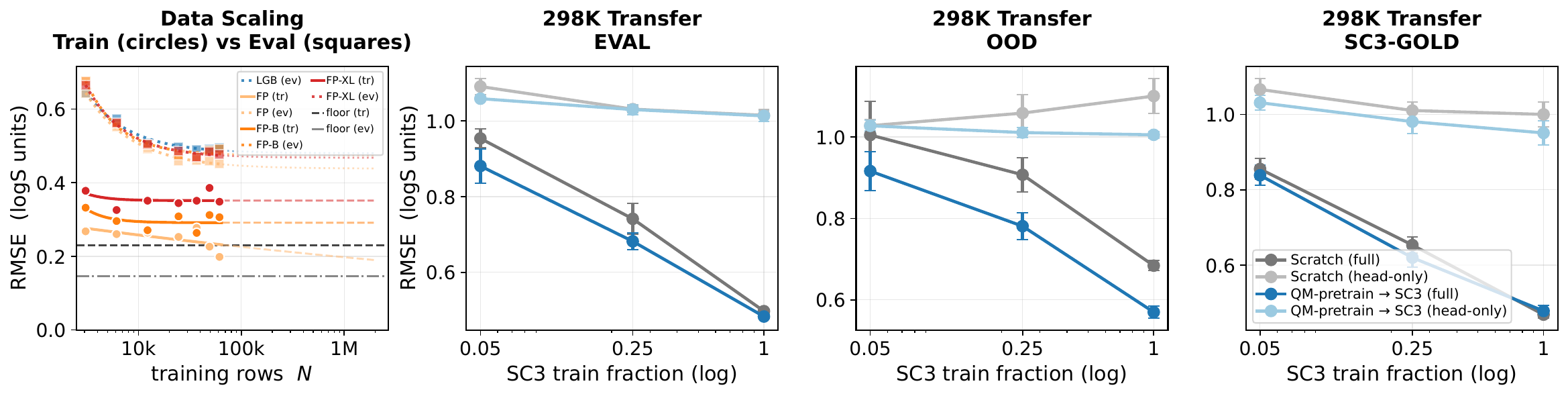}
\caption{\textbf{Data scaling and transferability.} (Left) Train vs.\ eval RMSE as a function of training set size. Train RMSE (solid lines, circles) and eval RMSE (dotted lines, squares) are overlaid for four models, along with their power-law fits and aleatoric floors. (Right panels) 298K-locked transfer learning evaluation on the eval, OOD, and sc3\_gold splits, comparing scratch models to those pretrained on QM data at various SC3 fine-tuning fractions.}
\label{fig:row2_scaling_transfer}
\end{figure*}

\input{sections/05c.transfer}

%% file: sections/05a.interpretations.tex
\subsection{What does the model actually learn?  Interpretability}
\label{sec:interpretability}

The benchmark numbers in §\ref{sec:baselines} answer \emph{how well} each
model predicts \logS{}; this section asks \emph{what} it has actually
learned.  We apply Tree-SHAP attribution across all seven LightGBM\,+\,featurizer
combinations and occlusion attribution to the GCN, without retraining any
model.  Full methodology, all supporting figures, and the per-solvent
breakdowns are in Appendix~§\ref{app:interp_full}.\\
Block-wise SHAP decomposition shows that the solute carries roughly
\numval{70}\,\% of total attribution mass across all representations,
with the solvent accounting for only {22}\,\%.
This share shifts when moving from in-distribution to OOD solvents:
every featurizer transfers \numval{3}--\numval{6} percentage points
from solute to solvent, which is exactly the behaviour expected of a
model that uses solvent identity as a coarse gate and falls back on it
harder when the solvent is unfamiliar.\\
The top global features reveal that
across all three descriptor-based representations, LightGBM
inherently recovers the axes of the Yalkowsky General Solubility
Equation---TPSA, BertzCT, and MolLogP on the solute side,
MaxPartialCharge as a solvent-polarity proxy, and normalised
temperature---without any chemistry prior
(Figure~\ref{fig:row1_interp}). The Abraham-only representation
recovers the canonical LSER ordering
$E > S \approx B \approx A > V$, and pairwise Tree-SHAP interaction
values reproduce the off-diagonal LSER cross-terms
($E\!\times\!E$, $A\!\times\!A$, $E\!\times\!B$) from rank 9 onward
(Appendix~Figure~\ref{fig:interp_interactions}).  This signals towards a thermodynamic structure emerging from data, rather than coefficient-fitting to a known equation.\\
The solvent clustering results explain the representation gap observed
in §\ref{sec:baselines}.  Descriptor-based models partition the 25
eval solvents into four chemistry-meaningful families---water, alkanes,
polar aprotic, and protic alcohols---with an Adjusted Rand
Index of \numval{0.21}--\numval{0.23} against the classical Snyder
taxonomy.  Fingerprint-based models produce a structurally-driven
grouping instead (ARI \numval{0.15}), placing n-hexane alongside
long-chain alcohols because they share carbon-chain Morgan bits rather
than functional class.  The model's internal representation fails on OOD solvents, not its capacity.\\
The GCN tells a complementary atom-level story
(Figure~\ref{fig:row1_interp}).  The BRICS fragment ranking is
interpretable throughout: every motif in the top~15 by aggregate
occlusion is a textbook solubility-relevant substructure (carboxylic
acid, phenol, nitrobenzene, piperazine), and no opaque fragment
appears in the top~20.  Signed atom-level maps for two representative
solutes confirm this: the carboxylic acid and nitro groups dominate
attribution in a nitroaromatic derivative, while the piperazine ring
drives prediction shifts in a diamine scaffold.  The GCN has learned
a chemistry-meaningful substructure ontology through message-passing
alone, without being told what these fragments are.


%% file: sections/05b.data_scaling.tex
\subsection{Data scaling vs.\ the aleatoric floor}
\label{sec:data_scaling}
Knowing whether a model fails because it has seen too little data, or
because its molecular representation cannot capture the relevant
chemistry, has direct implications for how to improve it.
To test this, we trained
\textsc{LightGBM} and \textsc{FastProp} MLPs on seven subsampled
fractions of \scthree-Train and tracked how test RMSE changed with
training-set size; the full protocol is in
Appendix~§\ref{app:data_scaling}.
We then fit a saturating power law, $\mathrm{RMSE}(N) = aN^{-b}+c$,
to each model's error curve.
The usefulness of this form is in its asymptote: as $N\to\infty$, the
curve converges to the constant $c$, so $c$ is the lowest error the
model could ever reach on a given split if given unlimited data, all
else held fixed.

Comparing $c$ against the split-specific aleatoric floor
$\varepsilon_A^{(s)}$---the irreducible measurement noise estimated
from repeated laboratory measurements
(Appendix~Table~\ref{tab:scaling_floors})---then tells us whether
greater data coverage could, even in principle, close the gap to the
noise limit.
Across all four models and evaluation splits, the fitted asymptotes
land $0.28$--$0.77$~log~S above their respective measurement floors,
and solving for the training-set size $N^\star$ at which the power
law would reach the floor yields $\infty$ in every case
(Appendix~Table~\ref{tab:scaling_extrapolation},
Appendix~Figure~\ref{fig:data_scaling_lawfits}).
The models do converge---most reach $95\%$ of their own asymptote
within $28\text{K}$--$217\text{K}$ training rows---but they converge
to an error level that remains well above what the noise floor permits.
The gap is therefore not a data-volume problem; it is a representation
problem, which motivates the transfer-learning study
of~§\ref{sec:transfer}.

%% file: sections/05c.transfer.tex
\subsection{Transfer learning: can adjacent chemistry guide solubility?}
\label{sec:transfer}
The data-scaling analysis of \S\ref{sec:data_scaling} established
that extra \scthree{} data does not, on its own, close the model gap
to the aleatoric floor.  An alternative is transferring inductive bias from
an adjacent task with far more data.  We test \textsc{CombiSolv-QM}
\citep{vermeire2021transfer}: ${\sim}10^6$ \textsc{cosmo-rs} solvation
free energies $\Delta G_{\text{solv}}$ at $298.15$~K, spanning
$11{,}029$ solutes and $284$ solvents, with the same
$(\text{solute}, \text{solvent})$ pair structure as \scthree{} and computational labels carrying essentially no experimental noise.
We pretrain a FastProp trunk on \textsc{CombiSolv-QM} and fine-tune
on fractions of \scthree{}-train, comparing against scratch baselines
under two setups: multi-temperature and $298$~K-locked (to isolate
the chemistry signal from temperature dependence).  Full experimental
design is in Appendix~\ref{app:transfer}.

\textsc{qm} pretraining shows improved performance in $9/9$ cells (multi-T) and $8/9$
(298~K-locked), never harming performance beyond seed variance.
The gain is largest when fine-tuning data is scarce: at $5$\%
of \scthree{}-train, the pretrained model matches scratch at
$25$--$100$\%---a $5$--$20{\times}$ data-efficiency improvement.
OOD gains are the most systematic ($-0.061, -0.033, -0.019$~\logS{}
at $5/25/100$\%), reflecting \textsc{CombiSolv-QM}'s $284$ solvents
covering the long-tail solvent space that \scthree{}-train
undersamples.  The $298$~K-locked setup yields a $6{\times}$ larger
OOD gain than multi-T at $100$\% data ($-0.114$ vs $-0.019$~\logS{}):
when the temperature axis is removed, the pretrained chemistry signal
is no longer overwritten by temperature re-learning during fine-tuning.
A frozen \textsc{qm} trunk with a $129$-parameter linear head
produces coherent predictions (\texttt{ood} \RMSE{} $= 0.98$) versus
\RMSE{} $\sim 10^7$ for a frozen random trunk---direct evidence that
\textsc{CombiSolv-QM} encodes a globally meaningful
solute--solvent representation.  The residual gap to LightGBM
($0.755$ vs $0.659$ on \texttt{sc3\_gold}) is consistent with
\S\ref{sec:data_scaling}: adjacent-task data moves the curve down
but does not close the architectural gap to tabular trees.

%% file: sections/06.conclusion.tex
\section{Conclusion}
We introduced \textsc{SC}$^3$, a multi-solvent solubility benchmark built on BIGSOLDB v2.1 with three core contributions: a reproducible curation pipeline yielding 101{,}535 measurements over 1{,}327 solutes and 206 solvents; a recalibrated aleatoric floor of $\varepsilon_{\text{aleatoric}} = 0.106$ log$S$---roughly $6\times$ tighter than the widely-cited 0.6--0.8 figure---exposing significant headroom above the true noise ceiling; and a standardised benchmarking protocol with nested consensus tiers, leakage-checked splits, and a metric suite headlined by PS-RMSE.\\
Our 31-model benchmark reveals that RDKit descriptors with tuned gradient-boosted trees remain the configuration to beat at $\approx 5 \times \varepsilon_{\text{aleatoric}}$ on Bronze PS-RMSE, with no deep or foundation-scale alternative closing this gap. Data-scaling analysis confirms this is a \emph{representation} bottleneck rather than a data-volume problem: power-law fits asymptote well above the aleatoric floor for every model and split, with $N^{\star}_{\varepsilon_A} = \infty$ in all cases. Pretraining on ${\sim}10^6$ quantum-chemistry solvation energies from COMBISOLV-QM partially addresses this gap, yielding systematic OOD gains and a 5--20$\times$ data-efficiency improvement at low fine-tuning fractions.\\
Together, these results reframe the state of solubility prediction: the field is not near a ceiling, but rather converging to a representation-limited plateau that more data alone cannot resolve. We release all splits, trained checkpoints, and analysis scripts to support future work targeting this gap. \textbf{Limitations:} the curation pipeline required manual interventions (unit-swap corrections, DOI-level outlier reviews) that are inherently subjective and may not generalise to future dataset releases; \textsc{SC}$^3$ occupies a crowded adjacent space (AQSOLDB, MIXTURESOLDB, concurrent multi-solvent efforts) where cross-benchmark comparison remains difficult; and while we clearly diagnose a representation bottleneck, we do not prescribe a concrete architectural path forward---whether the solution lies in equivariant geometry, explicit solvation-free-energy supervision, or a different molecular encoding remains open.

%% file: sections/appendix_data_curation.tex
\section{Data curation and aleatoric limit: extended material}
\label{app:data_curation}

This appendix consolidates all extended material supporting
\S\ref{sec:curation_aleatoric}: the detailed description of data curation steps, the per-solvent aleatoric decomposition
figure moved from the main text, threshold sensitivity for copycat
merging, full Apelblat fit-quality diagnostics, removed DOIs, and the
full canonicalization policy rationale.

\subsection{Raw data audit}
\label{app:raw_audit}

\textsc{BigSolDB}~v2.1 ships a companion table of experimentally measured solvent densities at
multiple temperatures alongside a main solubility dataset. Each row carries solute and solvent
SMILES, temperature, mole fraction~$x$, and a reported $\log_{10} S$~(mol/L) either derived
from~$x$ and solvent density or taken directly from the literature. Apart from one polymer-parsed
solute SMILES, all {1\,525} solute SMILES pass \texttt{RDKit} parsing.
 
\textbf{Missing \logS{} values.}
{3\,187} rows ({2.8\,\%}) carry \texttt{NaN} in the reported \logS{} column.
For the remaining {109\,278} rows we reconstructed \logS{} from~$x$ via the closed-form
relation
\begin{equation}\label{eq:closed_form}
    \log_{10} S = \log_{10}\!\left(\frac{x}{1-x}\cdot\frac{\rho(T)\,\cdot\,10^{3}}{M_w}\right),
\end{equation}
where $\rho(T) = aT + b$ is a per-solvent linear fit to the companion density data and $M_w$ is
obtained from the solvent SMILES via \texttt{RDKit}. Residuals against reported values are
negligible (median $|\mathrm{residual}| = 9.9\times 10^{-6}$, $P_{99} = 1.25\times 10^{-3}$);
only {8} rows exceeded a residual of $0.01$~log~S, all from one DOI ($\beta$-alanine in
methanol), which enters our bad-DOI list (Table~\ref{tab:bad_dois}). This agreement licenses
reuse of Eq.~\ref{eq:closed_form} to back-calculate \logS{} for \texttt{NaN} rows downstream
(step~W5, \S\ref{app:waterfall}).\\
\textbf{SMILES canonicalization.}
We canonicalize all solute and solvent SMILES with \texttt{RDKit}, preserving chirality and 
$E/Z$ geometry. We explicitly \emph{do not} enumerate tautomers, and the result is that all {1\,525} raw solute
SMILES remain distinct. Alternative approaches, like stripping stereochemistry or enumerating tautomers, were rejected because
the induced merges conflate compounds whose measured solubilities disagree by $5$--$20\times$ the aleatoric floor (Appendix~\ref{app:canonicalization}).\\
\textbf{Exact-duplicate rows.}
Grouping rows by (canonical solute, canonical solvent, $T$ rounded to $0.1$\,K, \logS{}) reveals
{86} such four-tuples reported identically by two or more DOIs---{172} rows across
{46} DOIs---an artifact of data duplication consumed when building independence groups for
the source-integrity analysis (\S\ref{app:source_integrity}).

\subsection{Source integrity and Thermodynamic Curve Fits}
\label{app:source_integrity}
To prevent duplicate measurements from deflating inter-lab statistics,
we adopt the following strategy to union DOIs into \emph{independence groups} via two-stage duplication detection and rank each group's reliability against peer consensus; unreliable groups feed the bad-DOI list removed at W1
(\S\ref{app:waterfall}).

\textbf{Stage A$'$ (bit-exact).} The {86} identical four-tuples from \S\ref{app:raw_audit} yield {21} inter-DOI unions.\\
\textbf{Stage B$'$ (interpolated).} Stage~A\textquotesingle\ misses rescaled or relabelled copies that are not bit-exact. After fitting per-group Apelblat / van't~Hoff curves (\S\ref{app:source_integrity}), we compute, for every preliminary-group pair sharing a (solute,\,solvent) pair, the MAE between their fitted curves on a uniform 1\,K grid over the intersection of their fit ranges (overlap $\geq 5$\,K required). Pairs whose weighted MAE falls below $\theta_{B'} = 0.01$~log~S are unioned. This merges {55} additional pairs, producing {1\,415} final independence groups (down from {1\,493} DOIs): {1\,348} singletons, {57} pairs, {8} triples, {2} quadruples. The $\theta_{B'} = 0.01$ cut sits in a sparse region of the pair-MAE distribution (Figure~\ref{fig:stages_bc}): only {55} of {373} tested pairs fall below it, while the bulk lies at $0.02$--$1$~log~S (which indicates genuine inter-lab disagreement).\\
\textbf{Stage C$'$ (reliability ranking).} Each group is scored by its mean absolute deviation from peer consensus across all shared (solute,\,solvent) grid points.  Of {399} scoreable groups, {316} are High-Reliability ($\leq 0.2$~log\,S) and {27} Low-Reliability ($\geq 0.6$~log\,S; Figure~\ref{fig:stages_bc}).

\textbf{Manual corrections.}
The {22} DOIs with mean inter-lab deviation $\geq 0.6\,\log S$ were
reviewed manually: four transcription bugs were corrected in place,
recovering {58} rows.  The remaining {5}, plus {3} from the
consensus-deviation audit and {1} residual outlier
(\S\ref{app:raw_audit}), form the {9} bad DOIs removed at W1
(Table~\ref{tab:bad_dois}).

\begin{table*}[t]
\centering\scriptsize
\renewcommand{\arraystretch}{0.85}
\caption{Left: manual corrections applied before the cleaning waterfall.
Right: Cleaning waterfall applied sequentially to the corrected, canonicalized data.}
\label{tab:waterfall_corrections}
\begin{minipage}[t]{0.42\linewidth}
\centering
\begin{tabular}{@{}c p{0.45\linewidth} l r@{}}
\toprule
Class & Correction & DOI & Rows \\
\midrule
C1 & paracetamol\,/\,water: four $x$ values replaced
   & \texttt{...113867} & 4 \\
C2 & $\log_{10} S$ shifted by $+1$
   & \texttt{...09.018}  & 20 \\
C2 & $\log_{10} S$ shifted by $+1$
   & \texttt{...06.011} & 10 \\
C3 & ethanol $\leftrightarrow$ ethyl acetate swapped, \logS{} re-derived
   & \texttt{...07.038}  & 24 \\
\bottomrule
\end{tabular}
\end{minipage}
\hfill
\begin{minipage}[t]{0.54\linewidth}
\centering
\begin{tabular}{@{}c l r r@{}}
\toprule
Step & Filter & Rows after & Removed \\
\midrule
    & Input (canonicalized, corrected)                             & 112\,465  & ---      \\
W1  & remove {9} bad DOIs                                   & 112\,019  & 446      \\
W2  & invalid / polymer solvent SMILES                             & 111\,705  & 314      \\
W3  & salts / mixtures (\texttt{.} in solute SMILES)               & 101\,663  & 10\,042  \\
W4  & $M_w > 1000$\,Da                                             & 101\,634  & 29       \\
W5  & \logS{} recovered for 2\,617 \texttt{NaN} rows              & 101\,620  & 14       \\
W6  & $\log_{10} S \notin [-15, 2]$                                & 101\,575  & 45       \\
W7  & intra-DOI near-duplicate dedupe                              & \textbf{101\,535} & 40 \\
\bottomrule
\end{tabular}
\end{minipage}
\end{table*}

\textbf{Thermodynamic Curve Fits.}
To compare laboratories that do not share exact measurement
temperatures, we fit a smooth thermodynamic model to every
(solute,\,solvent,\,group) triple: the Apelblat equation
($\log_{10} S = A + B/T + C \ln T$) for triples with $\geq 3$
temperature points, and a reduced van't~Hoff form
($\log_{10} S = A + B/T$) for exactly two points.  Single-point
triples (${\sim}\,4.5\,\%$) are excluded from aleatoric
interpolation, as are {16} fits with $R^2 < 0.80$ or
RMSE $> 0.30\,\log S$.  The remaining {11\,063} fits cover
$\geq 95\,\%$ of the cleaned data
(diagnostics in Figure~\ref{fig:apelblat}).

\subsection{Cleaning waterfall}
\label{app:waterfall}

After applying the manual corrections and canonicalization of
\S\ref{app:raw_audit}--\ref{app:source_integrity}, we apply a
seven-stage cleaning waterfall
(Table~\ref{tab:waterfall_corrections}).  Most steps are
self-explanatory from the table; we note that W3 removes salts and
multi-component solutes whose dissolution thermodynamics is
incomparable with that of neutral species, and W5 back-calculates
\logS{} for surviving \texttt{NaN} rows via
Eq.~\ref{eq:closed_form}, resolving $\rho(T)$ through a fallback
chain (companion density table $\to$ per-solvent linear fit $\to$
the \texttt{thermo} library).

\subsection{Per-solvent aleatoric decomposition}
\label{app:per_solvent_aleatoric}

\begin{figure}[h]
  \centering
  \includegraphics[width=0.9\linewidth]{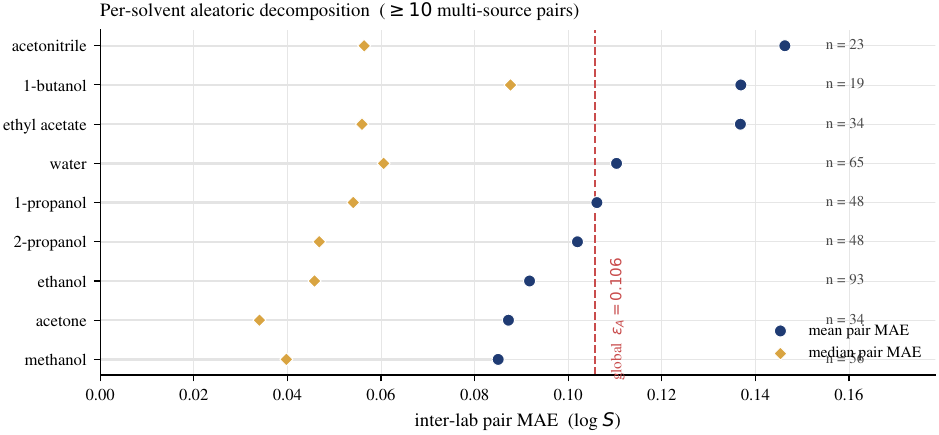}
  \caption{Per-solvent aleatoric decomposition (primary, LR-DOIs excluded,
    solvents with $\geq 10$ multi-source pairs). Lollipops are mean pair
    MAE; diamonds are median; $n$ is the multi-source pair count. Dashed
    vertical line is the global
    $\varepsilon_{\mathrm{aleatoric}} = 0.106$.}
  \label{fig:per_solvent_aleatoric}
\end{figure}

Figure~\ref{fig:per_solvent_aleatoric} shows the per-solvent breakdown
of the primary $\varepsilon_{\mathrm{aleatoric}}$ (LR-DOIs excluded)
for all solvents with $\geq 10$ multi-source pairs. The global
$\varepsilon_{\mathrm{aleatoric}} = 0.106$ (dashed vertical line) masks
substantial heterogeneity: DMF sits at $0.029$~log~S while water reaches
$0.110$~log~S, the latter driven by a thicker tail rather than a higher
typical disagreement (mean vs.\ median diverge most strongly for water).
This motivates reporting PS-RMSE as the headline metric (\S\ref{sec:metrics}).

\subsection{Threshold sensitivity for copycat merging ($\theta_{B'}$)}
\label{app:threshold_sensitivity}

\begin{figure}[h]
  \centering
  \includegraphics[width=0.9\linewidth]{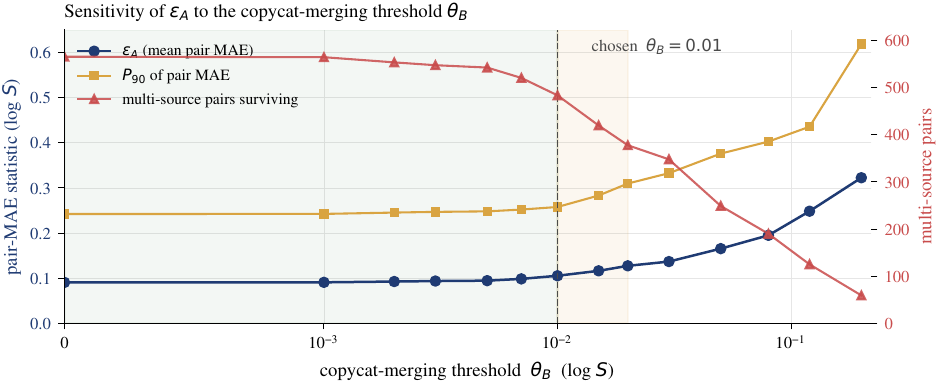}
  \caption{Sensitivity of $\varepsilon_{\mathrm{aleatoric}}$ and the
    multi-source pair count to the copycat-merging threshold $\theta_B$.
    The chosen value $\theta_B = 0.01$ (dashed line) sits at the upper
    edge of the shaded ``safe'' region, where each merged pair is a
    genuine copy. The shaded ``cliff'' region $[0.01, 0.02]$ loses
    \numval{106} multi-source pairs---more than the entire loss incurred
    between $0$ and $0.01$.}
  \label{fig:threshold_sweep}
\end{figure}

The threshold $\theta_{B'}$ controls a trade-off: too small a value
leaves obvious copies in the multi-source pool where they deflate
$\varepsilon_{\mathrm{aleatoric}}$ artificially; too large a value
merges genuinely agreeing labs, shrinks the multi-source pool, and
inflates $\varepsilon_{\mathrm{aleatoric}}$ among the smaller remaining
pairs. Figure~\ref{fig:threshold_sweep} shows the empirical curve.
Below $\theta = 0.007$ very few additional unions are made: the
multi-source pair count barely decreases (from \numval{566} at $\theta =
0$ to \numval{521} at $\theta = 0.007$, 8\,\% loss) and
$\varepsilon_{\mathrm{aleatoric}}$ rises only 8\,\% (from \numval{0.091}
to \numval{0.098}). Between $\theta = 0.010$ and $\theta = 0.020$ there
is a cliff in the pair count: from \numval{484} pairs to \numval{378}
pairs, a loss of \numval{106} pairs over a 0.01-log-S interval (more
than the entire loss for $\theta \in [0, 0.01]$). At the same time
$\varepsilon_{\mathrm{aleatoric}}$ jumps by 21\,\%, from \numval{0.105}
to \numval{0.127}. Beyond $\theta = 0.05$ the multi-source pool has more
than halved. We therefore adopt $\theta = 0.01$ as the conservative
upper edge of the ``copycats-only'' region: it merges the detectable
copies while preserving the largest possible multi-source pool for
downstream analysis.

\subsection{Apelblat fit-quality diagnostics}
\label{app:apelblat_diagnostics}

\begin{figure}[h]
  \centering
  \includegraphics[width=\linewidth]{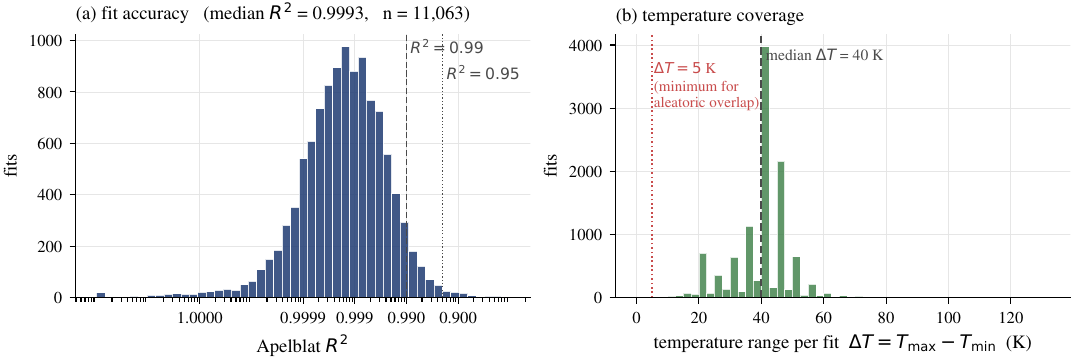}
  \caption{Apelblat fit quality. \textbf{(a)}~$R^2$ distribution on a
    reversed log scale ($1 - R^2$ axis). Most fits have $R^2 > 0.99$.
    \textbf{(b)}~Temperature coverage $\Delta T$ per fit. Most fits
    cover more than the 5~K minimum needed for aleatoric interpolation;
    the median fit spans \numval{30}~K.}
  \label{fig:apelblat}
\end{figure}

Figure~\ref{fig:apelblat}(a) shows the fit-quality distribution: median
$R^2 = 0.9993$, \numval{99.3\,\%} of fits reach $R^2 \geq 0.95$,
\numval{94.9\,\%} reach $R^2 \geq 0.99$. Thermodynamic monotonicity is
similarly strong: \numval{98.5\,\%} of triples are strictly
monotone-increasing in temperature and only \numval{1} triple shows a
single-step drop exceeding $1$~log~S. Figure~\ref{fig:apelblat}(b)
shows the distribution of per-fit temperature coverage $\Delta T =
T_{\max} - T_{\min}$. The median fit spans \numval{30}~K and only a
small minority of fits cover less than the 5~K minimum required for
cross-group interpolation, confirming that the
Apelblat framework supports aleatoric analysis on the majority of the
multi-source pool without extrapolation.

\subsection{DOIs removed at stage W1}
\label{app:bad_dois}

\begin{table}[h]
\centering\footnotesize
\caption{Nine DOIs removed at W1 of the cleaning waterfall
(\S\ref{app:waterfall}).}
\label{tab:bad_dois}
\begin{tabular}{@{}l l@{}}
\toprule
Source of flag & DOI \\
\midrule
\multirow{5}{*}{Manual review}
  & \texttt{10.1016/j.fluid.2011.09.033} \\
  & \texttt{10.1021/jced.9b00728} \\
  & \texttt{10.1021/jced.4c00179} \\
  & \texttt{10.1021/jced.6b00009} \\
  & \texttt{10.1016/j.molliq.2022.119759} \\
\midrule
\multirow{3}{*}{Consensus-deviation audit (\S\ref{app:source_integrity})}
  & \texttt{10.1021/je4000718} \\
  & \texttt{10.1016/s1004-9541(08)60201-3} \\
  & \texttt{10.1016/j.molliq.2016.11.036} \\
\midrule
\logS{}-residual outlier
  & \texttt{10.1016/j.molliq.2017.02.075} \\
\bottomrule
\end{tabular}
\end{table}

\subsection{Canonicalization policy: full rationale}
\label{app:canonicalization}

This appendix records the alternative SMILES-canonicalization policies
we considered, the empirical evidence against the merging policies, and
the chemical interpretation of that evidence. The main text adopts
\emph{Option~D} (plain RDKit canonical with all stereochemistry
preserved); the material below explains why.

\subsubsection{Candidate policies}

\begin{table}[h]
\centering\small
\caption{Candidate canonicalization policies for solute SMILES and the
resulting unique-solute count when applied to the \numval{1\,525} raw
solute SMILES in \textsc{BigSolDB}~v2.1.}
\label{tab:canonicalization}
\begin{tabular}{@{}clr@{}}
\toprule
Option & Definition & Unique solutes \\
\midrule
A & tautomer-enumerate $+$ strip \texttt{@/@@} $+$ strip \texttt{/\textbackslash} & 1\,506 \\
B & no tautomer $+$ strip \texttt{@/@@} $+$ strip \texttt{/\textbackslash}        & 1\,508 \\
C & no tautomer $+$ strip \texttt{@/@@}, keep \texttt{/\textbackslash}             & 1\,510 \\
\textbf{D} & plain canonical, \texttt{isomericSmiles=True} (keep all stereo)      & \textbf{1\,525} \\
\bottomrule
\end{tabular}
\end{table}

Options A--C all merge distinct compounds. Two concrete failures of
stereo stripping illustrate the cost: \emph{fumaric} acid (trans,
\verb|O=C(O)/C=C/C(=O)O|) and \emph{maleic} acid (cis,
\verb|O=C(O)/C=C\C(=O)O|), which differ in aqueous solubility by
roughly $100\times$, become identical after removing
\texttt{/}/\texttt{\textbackslash}. A concrete failure of tautomer
enumeration: a bicyclic norbornene anhydride is silently aromatized by
the RDKit tautomer enumerator into a structurally nonsensical ``aromatic
diol''.

\subsubsection{Empirical audit of Option-C merges}

To choose between the remaining options we audit every Option-C merge
group: for each group we find all $(\text{solvent}, T)$ cells in which
$\geq 2$ raw-SMILES members report measurements and compute $|\Delta
\log_{10} S|$ at matched conditions. If enantiomers genuinely had
identical solubility in achiral solvents, these deltas should lie at or
below the aleatoric floor ($\sim 0.1$~log~S; \S\ref{sec:aleatoric}). We
find the opposite: \emph{10 of 15} Option-C merge groups disagree by a
median of \numval{0.29}--\numval{1.31}~log~S across 3--117 overlap pairs
(Table~\ref{tab:stereo_audit}), i.e.\ $5$--$20\times$ the aleatoric
floor.

\begin{table}[h]
  \centering\small
  \caption{Empirical audit of Option-C stereo-merge groups. Median / max
    $|\Delta \log_{10} S|$ across all $(\mathrm{solvent}, T)$ cells
    where $\geq 2$ raw-SMILES members of the merged group have
    measurements. The four ``silent'' merges (bottom) have no overlap
    cells and cannot be empirically tested.}
  \label{tab:stereo_audit}
  \begin{tabular}{@{}lrrrr@{}}
  \toprule
  Merge group (common name) & total rows & overlap pairs & median $|\Delta|$ & max $|\Delta|$ \\
  \midrule
  Brassinolide (2 stereoisomers)  & 146 &  36 & \textbf{1.31} & 2.01 \\
  D / L-psicose                   & 179 &  36 & 0.61          & 0.89 \\
  D / L-malic acid                & 141 &  21 & 0.52          & 0.99 \\
  D / L-tryptophan                & 229 &  84 & 0.43          & 0.97 \\
  D / L-norbornene anhydride      & 245 & 117 & 0.40          & 0.56 \\
  L-leucine / mixed               & 126 &   8 & 0.39          & 0.45 \\
  D / L-tyrosine                  & 123 &  16 & 0.37          & 1.07 \\
  N-acetyl-methionine D / L       & 188 &  45 & 0.37          & 0.73 \\
  Ibuprofen (S vs mixed)          & 106 &   1 & 0.29          & 0.29 \\
  \midrule
  L-isoleucine (silent)           & 126 &   0 & ---           & ---  \\
  Ofloxacin (silent)              & 124 &   0 & ---           & ---  \\
  Naproxen (silent)               & 115 &   0 & ---           & ---  \\
  Tartaric acid (silent)          &  49 &   0 & ---           & ---  \\
  \bottomrule
  \end{tabular}
\end{table}

\subsubsection{Chemical interpretation}

The main chemically interpretable explanation for the observed
disagreement is that enantiomeric compounds can be crystallized in
different polymorphic modifications, for which solubility differences
are commonly observed. Other potential causes are
lab~$\times$~chirality-label correlation in measurement protocol, or
silent mislabelling of racemate as L (or vice versa) at the
\textsc{BigSolDB} extraction stage. We cannot disambiguate these from
the data, but \emph{any} of the three implies that merging averages away
a $\sim 0.4$-log-unit signal we cannot recover. Adopting Option~D
(stereo-preserving canonical SMILES) keeps that signal in the labels;
chirality-blind featurizations then pay the cost in model error rather
than hiding it inside the label.

\subsection{Anti-leakage verification}
\label{app:antileakage}

\begin{table}[]
\centering\small
\caption{Anti-leakage verification.  Every cell reports the count of
(solute, solvent) pair overlaps between two splits.  Zero everywhere
except the two off-diagonal entries that are expected to share solutes
but never pairs (shaded).  All checks pass.}
\label{tab:antileakage}
\begin{tabular}{@{}lcccccc@{}}
\toprule
          & Train & Eval & OOD & Gold & Silver & Bronze \\
\midrule
Train     & ---   & 0    & 0   & 0    & 0      & 0      \\
Eval      & 0     & ---  & 0   & 0    & 0      & 0      \\
OOD       & 0     & 0    & --- & 0    & 0      & 0      \\
Gold      & 0     & 0    & 0   & ---  & nested & nested \\
Silver    & 0     & 0    & 0   & nested & ---  & nested \\
Bronze    & 0     & 0    & 0   & nested & nested & ---   \\
\bottomrule
\end{tabular}
\end{table}

\subsection{Per-solvent log~$S$ distributions}
\label{app:solvent_distributions}

\begin{figure}[ht]
\centering
\includegraphics[width=\linewidth]{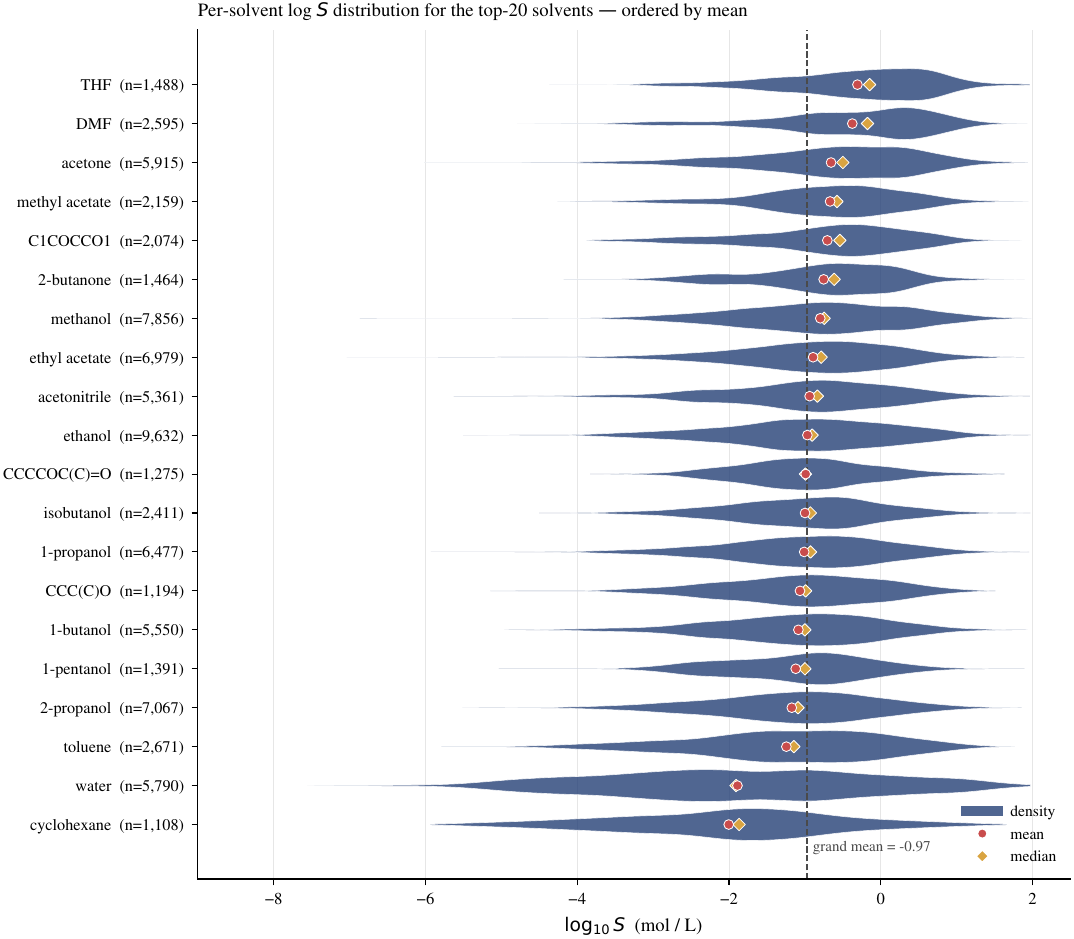}
\caption{Per-solvent log~$S$ distribution (violin) for the top-20
  solvents by row count, ordered by per-solvent mean.  Red marker =
  mean, gold diamond = median, dashed line = grand mean.  The top-to-
  bottom mean span is $\sim 7$ log~$S$; \emph{any} predictor that
  correctly identifies the solvent absorbs most of that variance for
  free.}
\label{fig:solvent_dist}
\end{figure}

%% file: sections/appendix_interpretations.tex
\section{Interpretability: full methodology and extended results}
\label{app:interp_full}

This appendix contains the full methodology, all supporting figures, and
extended analysis for the interpretability study summarised in
§\ref{sec:interpretability} of the main paper.

We hold the model fixed (LightGBM with the tuned \texttt{lgb\_rdkit}
hyperparameters of §\ref{sec:ablations}) and rerun it under seven
different molecular representations -- the same seven studied in the
Representation ablation -- so the only thing changing across runs is how
chemistry enters the input.  For each (model, representation) we compute
exact Tree-SHAP values~\cite{lundberg2017unified} on the full
\texttt{eval} and \texttt{ood} splits; then we slice those attributions by
feature, by solvent, and by feature pair to expose \emph{which} chemistry the
model has internalised.  In addition, we re-use the trained dual-encoder GCN
(§\ref{sec:baselines}) and attribute its predictions back to atoms and BRICS
fragments via occlusion analysis.  No model is retrained for this study;
everything is a post-hoc decomposition of predictions already reported in the
benchmark table.

Sanity check: per-featurizer \RMSE{} on \texttt{eval}/\texttt{ood} exactly
reproduces the Representation-ablation numbers (\eg{} RDKit \texttt{eval}
\RMSE{}~$=$~\numval{0.493}, \texttt{ood}~$=$~\numval{0.605};
GCN \texttt{eval}~$=$~\numval{0.608}), so every importance plot below is
anchored to a known-good predictor.  The full set of per-feature-set CSVs and
per-solvent JSONs is released.

\subsection{Where the signal comes from}
\label{app:interp_blockwise}

\begin{figure}[t]
\centering
\includegraphics[width=\linewidth]{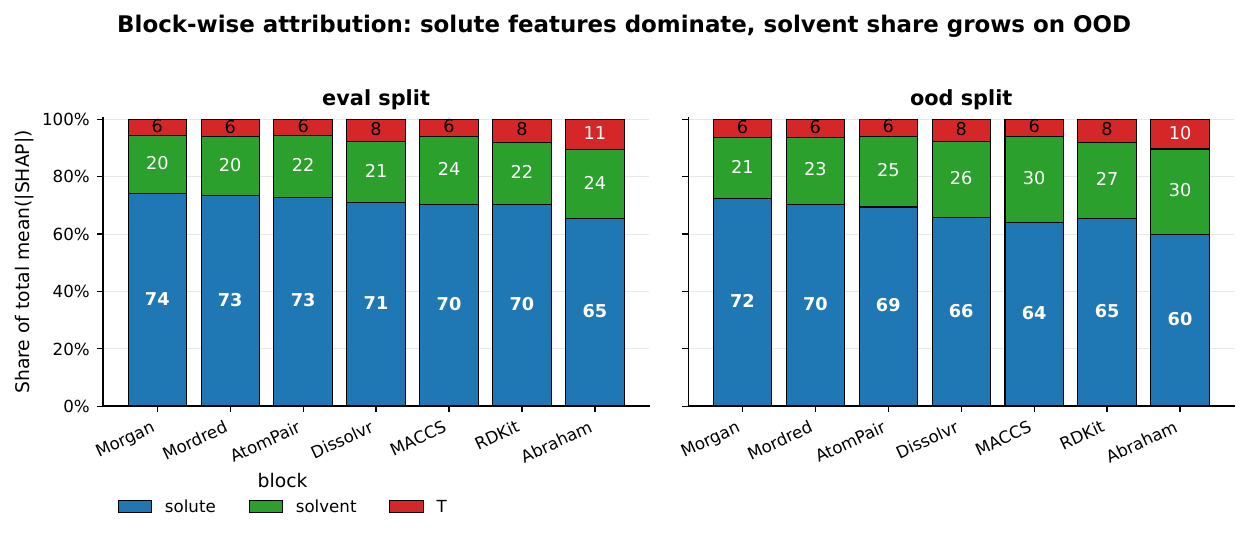}
\caption{\textbf{Block-wise SHAP attribution.}  Stacked bars show the
fraction of total mean(|SHAP|) coming from the solute, solvent, and
temperature blocks for each of the seven LightGBM\,$+$\,featurizer runs, on
\texttt{eval} (left) and \texttt{ood} (right).  The \emph{solute} dominates
universally (\numval{60}--\numval{74}\,\%) and the share is remarkably stable
across representations.  Going from \texttt{eval} to \texttt{ood} consistently
shifts mass from solute to solvent (\(+3\)--\(+6\,\)pp): when the solvent set
is unfamiliar the model leans more on solvent features.}
\label{fig:interp_blockwise}
\end{figure}

The first decomposition we report is also the bluntest.  Every featurizer
concatenates [\,solute~features\,$\,|\,$\,solvent~features\,$\,|\,$\,4
temperature features\,] into one input vector, so we can sum mean(|SHAP|)
within each block and ask: of the model's total attribution mass, how much is
the solute worth, how much is the solvent worth, how much is temperature worth?
Figure~\ref{fig:interp_blockwise} answers this for all seven representations
on both splits.

Three observations stand out.

First, the solute carries roughly \numval{70}\,\% of the total attribution mass
on \texttt{eval} and the solvent only \numval{20}\,\%, almost independently of
representation.  This seems counter-intuitive (although explainable as solute has far more chemical descriptors due to its higher molecular complexity) given the SC$^{3}$ premise that
solubility is a function of the \emph{pair}, not of either component alone, and
it reframes what tabular models trained on \scthree{} are actually optimising:
most of the loss is reduced by knowing the solute, and solvent identity is a
comparatively modest correction term.

Second, the temperature block consumes \numval{6}--\numval{11}\,\% of the
attribution mass.  Abraham-only is the outlier (\numval{11}\,\% on
\texttt{eval}) because with only 6 chemistry features per molecule the model has
less to spend, so the four temperature features become relatively more
important.

Third, every featurizer shifts \(\approx +3\)--\(+6\) percentage points from
solute to solvent when moving \texttt{eval}\(\to\)\texttt{ood}
(MACCS: \numval{24}\,\%\(\to\)\numval{30}\,\%;  Abraham:
\numval{24}\,\%\(\to\)\numval{30}\,\%).  This is exactly the behaviour a
well-generalising model should show: when the test solvents differ from the
training solvents, the model invests more attribution in the solvent features it
can still recognise.  Solvent identity is the ground-truth nuisance variable;
the model's response to OOD is to lean on it harder.

\subsection{Top global features and the General Solubility Equation}
\label{app:interp_global}

Drilling one level deeper, Figure~\ref{fig:row1_interp} (main paper)
shows the top-12 features for the three chemistry-readable representations
(RDKit, Dissolvr, Abraham-only).  We omit the fingerprint variants because
their top features are individual hashed bits without a direct chemistry
interpretation.

The headline result is that LightGBM, given no prior chemistry knowledge,
independently rediscovers \emph{the same axes} that classical solubility
equations were built around.  The Yalkowsky General Solubility
Equation~\cite{yalkowsky1980solubility} is
\(\logS \approx 0.5 - \log P - 0.01\,(T_m - 25)\), a linear function of just
two solute properties; SHAP says the LightGBM model is fundamentally doing the
same thing -- only with TPSA and BertzCT included as additional polarity and
complexity correction terms, plus a temperature-from-data correction.

A quantitative aside: classical GSE puts the melting-point term
(\texttt{pred\_Tm}) at coefficient \(-0.01\) and gives it large predictive
weight.  In our SHAP ranking on Dissolvr (which contains both \texttt{MolLogP}
and the Joback \texttt{pred\_Tm} proxy), \texttt{pred\_Tm} only reaches global
mean(|SHAP|) \(\approx\) \numval{0.029}, an order of magnitude below
\texttt{TPSA} (\numval{0.240}) and \texttt{BertzCT} (\numval{0.198}).  Our
data say polarity dominates crystal packing for \scthree{}-style liquid
solubility predictions much more than Yalkowsky's coefficients suggest.

The Abraham-only panel is the cleanest test of the LSER analogue: running
LightGBM on just six features per molecule
(\(A, B, S, E, V, T_m\)) gives a top ranking of
\(E_{\mathrm{solute}} > S_{\mathrm{solute}} > B_{\mathrm{solute}}
\approx A_{\mathrm{solute}} > V_{\mathrm{solute}}\), matching the canonical
LSER coefficient ordering for aqueous and organic
solubility~\cite{abraham1993scales}.  The first solvent-side feature is
\(B_{\mathrm{solv}}\) (H-bond basicity), at rank 6, which is the correct LSER
axis to rank first on the solvent side as well.

The magnitude plots above answer the question ``what matters?'' but not the
complementary question ``in which direction?''  To get a directional readout
we compute, for every feature, Spearman's \(\rho\) between the raw feature
value \(x_j\) and its SHAP contribution \(\phi_j(x)\) over the \texttt{eval}
split.

\begin{figure}[t]
\centering
\includegraphics[width=\linewidth]{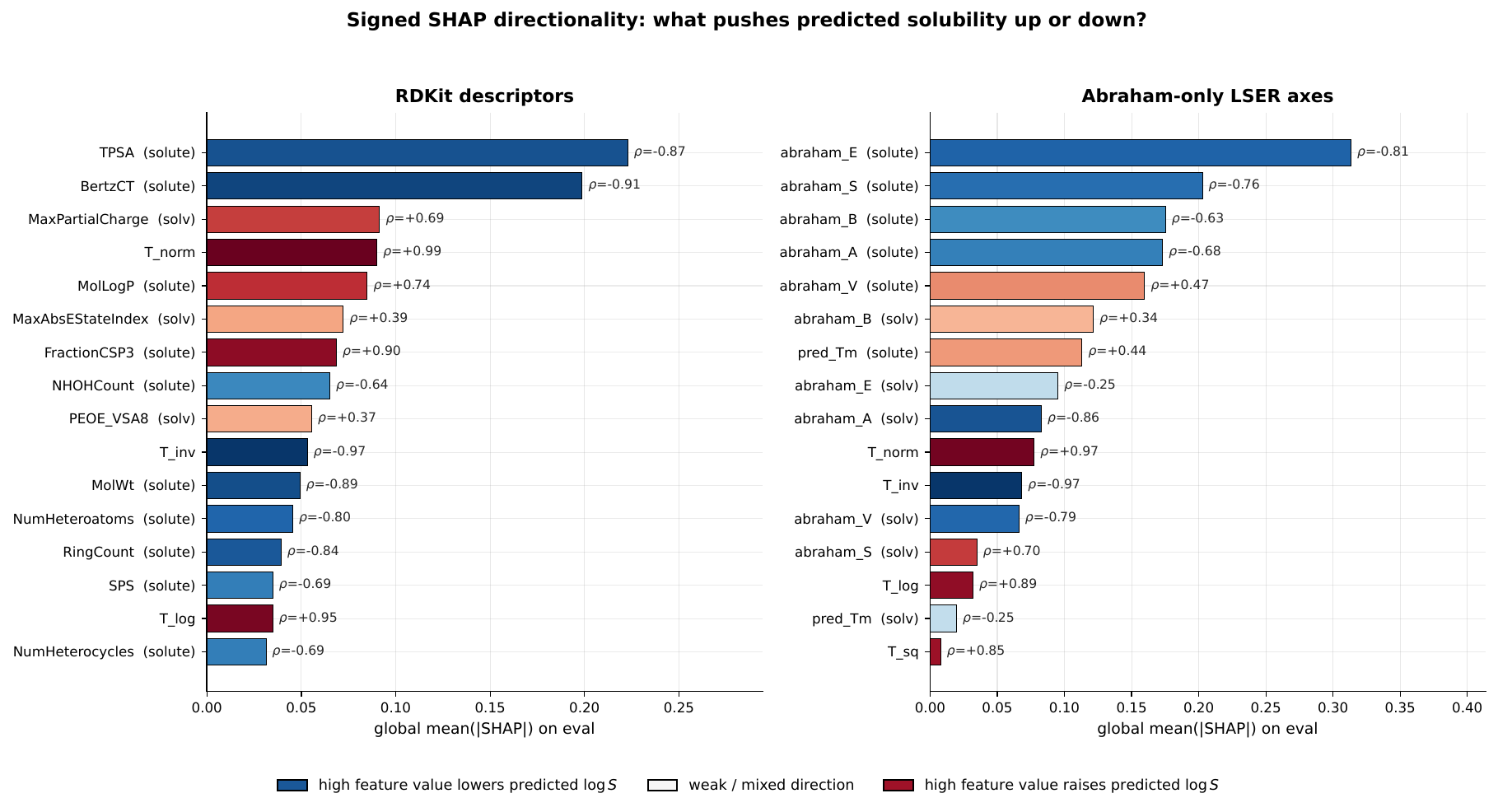}
\caption{\textbf{Signed SHAP directionality.}  Bar length is the usual global
mean(|SHAP|), so the ranking is unchanged from
Figure~\ref{fig:row1_interp}.  Bar colour and the printed Spearman
\(\rho\) show direction: blue features have high values that lower predicted
\logS{}, red features have high values that raise predicted \logS{}.  The
descriptor panel says the strongest solute features -- \texttt{TPSA},
\texttt{BertzCT}, \texttt{MolWt}, \texttt{RingCount}, heteroatom counts --
mostly push predictions down when large, while temperature and solvent
partial-charge features push predictions up.  The Abraham-only panel makes the
LSER story explicit: solute \(E,S,B,A\) lower predicted solubility when large,
whereas increasing temperature raises it.}
\label{fig:interp_signed_shap}
\end{figure}

Two details are worth noting.  First, high \texttt{T\_norm} has
\(\rho\approx +0.99\) and high \texttt{T\_inv} has \(\rho\approx -0.97\),
exactly the expected thermodynamic monotonicity: the model has learned that
higher temperature usually increases solubility.  Second, the solute descriptor
directions are not simply ``more polar means more soluble''.  On the
multi-solvent aggregate, large TPSA and large Abraham \(E/S/B/A\) values tend
to lower the global prediction.  This is not contradictory to aqueous
intuition: the dataset is dominated by organic solvents, and the signed SHAP
score is a conditional association in the fitted model, not a one-solvent
causal law.  The per-solvent analysis below is therefore essential; the global
signed plot is a direction-of-effect summary, not a universal chemical rule.

\subsection{Per-solvent profiles}
\label{app:interp_per_solvent}

\begin{figure}[t]
\centering
\includegraphics[width=\linewidth]{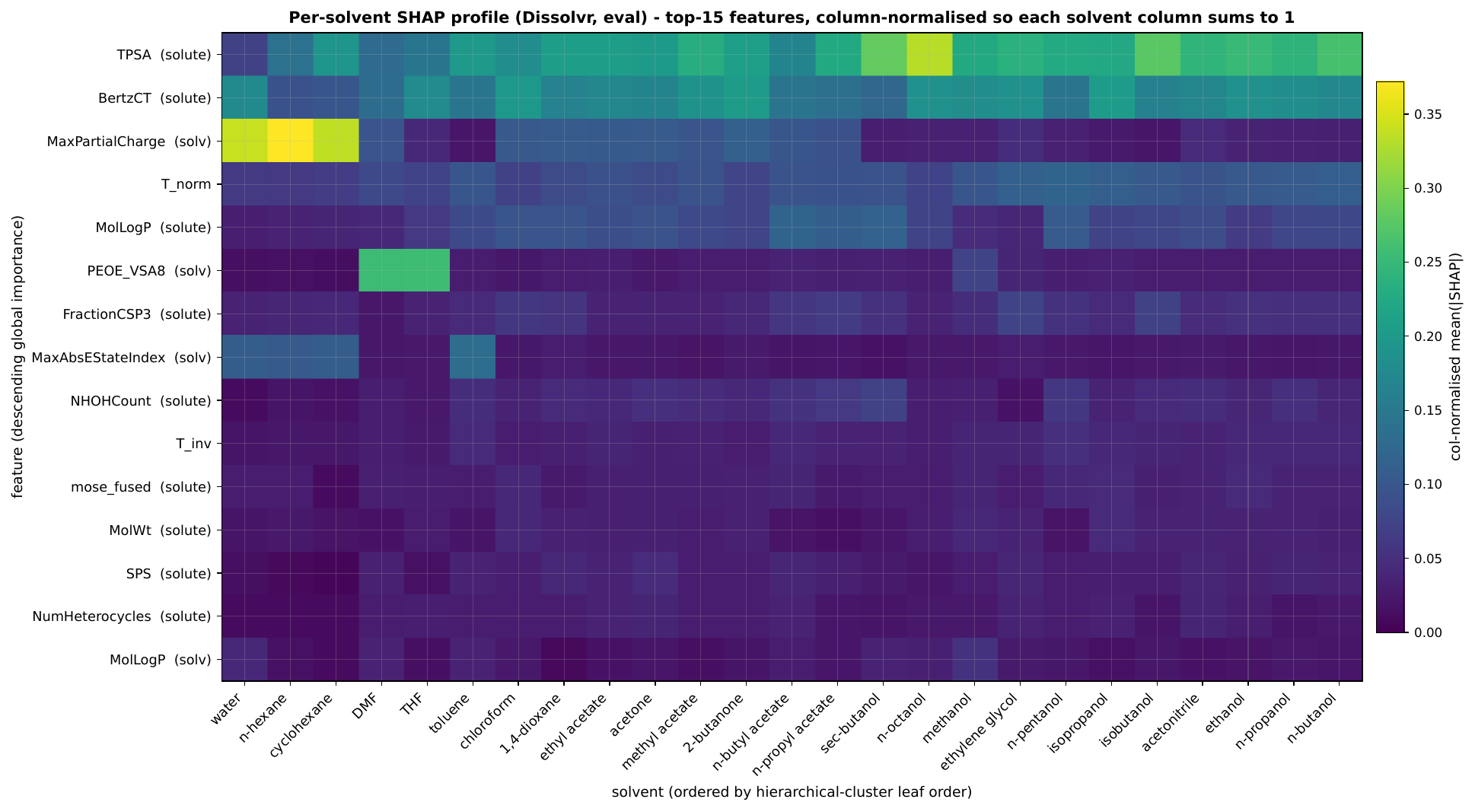}
\caption{\textbf{Per-solvent SHAP profile (Dissolvr, eval).}  Rows are the
top-15 globally-important features (descending); columns are the 25
\texttt{eval} solvents, ordered by hierarchical clustering on the column
vectors.  Cells are column-normalised mean(|SHAP|), so darker \(=\) less
important relative to the rest of that solvent's profile.  Two qualitative
patterns are visible: (i) for typical organic solvents the model relies on
\texttt{TPSA} and \texttt{BertzCT} of the \emph{solute}; (ii) for the extreme
solvents (water, n-hexane, cyclohexane) it switches to
\texttt{MaxPartialCharge} and \texttt{MaxAbsEStateIndex} of the
\emph{solvent}.  A third specialised pattern shows up for DMF and THF, where
\texttt{PEOE\_VSA8} of the solvent dominates.}
\label{fig:interp_per_solvent_heatmap}
\end{figure}

We can also slice the SHAP attribution by solvent.  For each of the top-25
in-distribution solvents (the \texttt{eval} solvent set), we restrict to the
rows where that solvent is the medium and compute mean(|SHAP|) per feature.
Figure~\ref{fig:interp_per_solvent_heatmap} shows the result for Dissolvr;
each column sums to one, so the heatmap reads as ``which feature does the model
lean on, given this solvent''.

Three distinct per-solvent regimes appear: ``alcohols and esters'' (the broad
band on the right where \texttt{TPSA(solute)} dominates), ``alkanes \(+\)
water'' (the leftmost columns where \texttt{MaxPartialCharge(solv)} dominates),
and ``aprotic-aromatic'' (DMF, THF, with \texttt{PEOE\_VSA8(solv)} dominant).
Per-solvent top-3 feature lists are summarized in
Table~\ref{tab:interp_anchor_solvents} for the five anchor solvents most often
used as references in the solubility literature.

\begin{table}[t]
\centering\small
\caption{Per-solvent top-3 SHAP features (RDKit, eval).  ``solv'' prefix
denotes solvent-side features; otherwise the feature is on the solute.  N is
the number of \texttt{eval} rows for that solvent.}
\label{tab:interp_anchor_solvents}
\renewcommand{\arraystretch}{1.10}
\begin{tabular}{@{}llrl@{}}
\toprule
Solvent      & N    & rank 1 (mean\,|SHAP|)            & ranks 2-3 \\
\midrule
\textbf{water}        & \numval{397} & \texttt{solv\_MaxPartialCharge}  (\numval{0.50}) & \texttt{solv\_MaxAbsEStateIndex}, \texttt{solute\_BertzCT} \\
\textbf{n-hexane}     & \numval{120} & \texttt{solv\_MaxPartialCharge}  (\numval{0.53}) & \texttt{solv\_MaxAbsEStateIndex}, \texttt{solute\_TPSA} \\
\textbf{DMF}          & \numval{263} & \texttt{solv\_PEOE\_VSA8}        (\numval{0.34}) & \texttt{solute\_BertzCT}, \texttt{solute\_TPSA} \\
\textbf{ethanol}      & \numval{782} & \texttt{solute\_TPSA}             (\numval{0.24}) & \texttt{solute\_BertzCT}, \texttt{T\_norm} \\
\textbf{methanol}     & \numval{594} & \texttt{solute\_TPSA}             (\numval{0.23}) & \texttt{solute\_BertzCT}, \texttt{solv\_PEOE\_VSA8} \\
\textbf{toluene}      & \numval{222} & \texttt{solute\_TPSA}             (\numval{0.24}) & \texttt{solv\_MaxAbsEStateIndex}, \texttt{solute\_BertzCT} \\
\textbf{chloroform}   & \numval{ 80} & \texttt{solute\_BertzCT}          (\numval{0.24}) & \texttt{solute\_TPSA}, \texttt{solute\_MolLogP} \\
\textbf{acetonitrile} & \numval{456} & \texttt{solute\_TPSA}             (\numval{0.24}) & \texttt{solute\_BertzCT}, \texttt{solute\_MolLogP} \\
\bottomrule
\end{tabular}
\end{table}

The pattern is consistent: in 22 of 25 \texttt{eval} solvents two of the top-3
features are solute features.  Only for water and the two alkanes (n-hexane,
cyclohexane) do solvent-side features take the top-2 slots, as these solvents are by a wide margin "guiding" the solubility, being one of the most (water) and least (alkanes) polar across the database. This suggests the
model uses solvent-side features as a coarse \emph{gate}: ``decide what kind of
solvent we are in, then evaluate the solute against that gate's specialised
sub-model''.  The number of distinct gates is small -- about four, matching the
four solvent families recovered by the clustering analysis of
§\ref{app:interp_clustering}.

\subsection{Solvents the model treats similarly}
\label{app:interp_clustering}

Each solvent's SHAP profile vector (§\ref{app:interp_per_solvent}) defines a
per-solvent fingerprint in feature-importance space.  Two solvents with similar
fingerprints are, by construction, treated similarly by the model: it weights
the same features when predicting solubility in either.  We compute the cosine
similarity between every pair of solvent fingerprints (using the top-50
globally-important features as the vector basis) and run hierarchical
clustering with average linkage.

\begin{figure}[t]
\centering
\includegraphics[width=\linewidth]{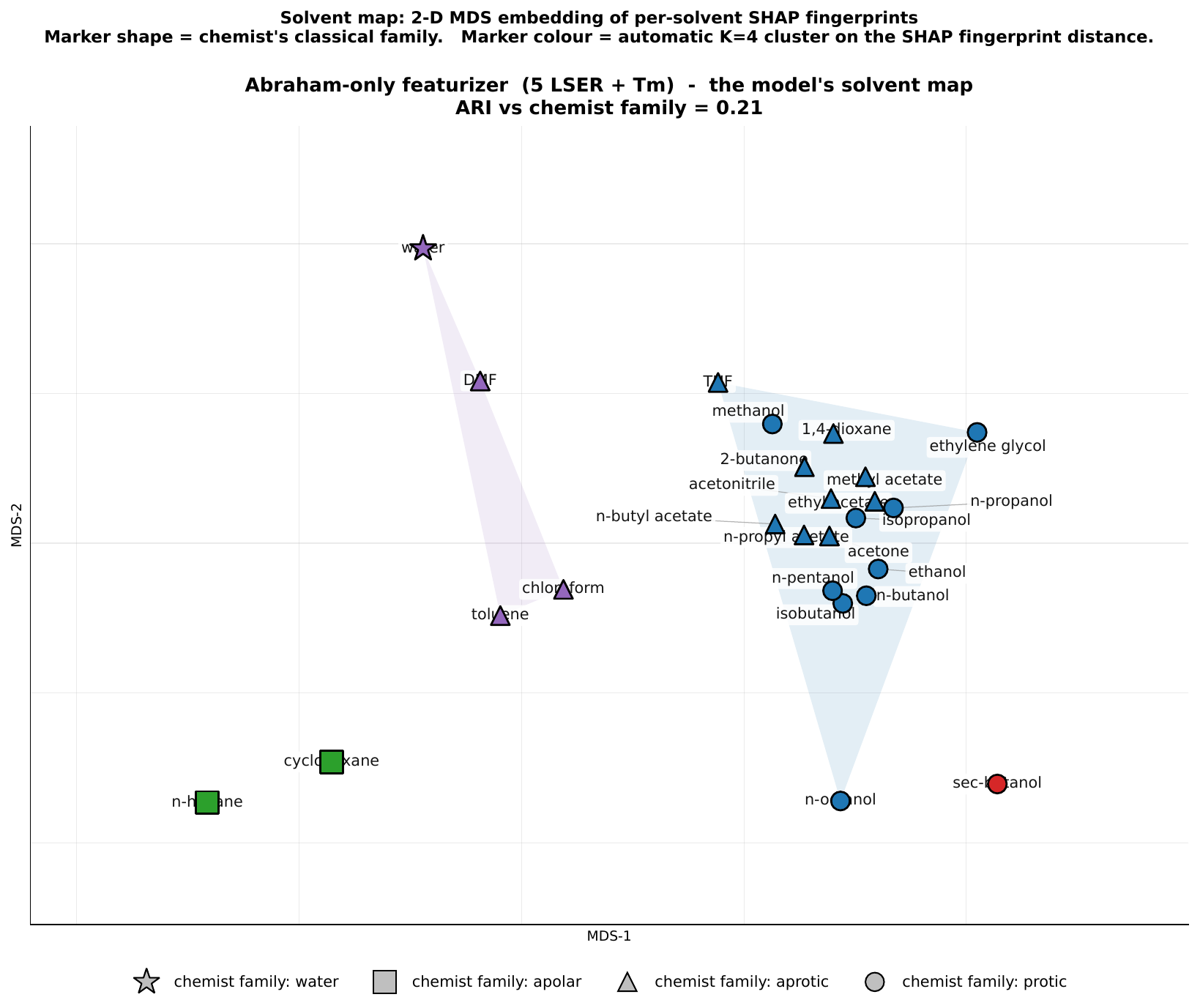}
\caption{\textbf{Solvent map under the Abraham-only featurizer.}  Each point
is one of the 25 \texttt{eval} solvents at its 2-D MDS coordinates derived
from the per-solvent SHAP fingerprint distance \(1 - \cos(\cdot)\).
\textbf{Marker shape} encodes the chemist's classical Snyder family
(\(\star\)~water, \(\blacksquare\)~apolar alkane, \(\blacktriangle\)~aprotic,
\(\bullet\)~protic alcohol).  \textbf{Marker colour} encodes the model's
automatic cluster (K=4 average-linkage hierarchical, on the same distance).
Same colour \(=\) the model treats those solvents the same way; same shape
\(=\) chemists treat them the same way.  ARI vs the chemist
labelling \(=\) \numval{0.21}.}
\label{fig:interp_solvent_map}
\end{figure}

\begin{figure}[t]
\centering
\includegraphics[width=\linewidth]{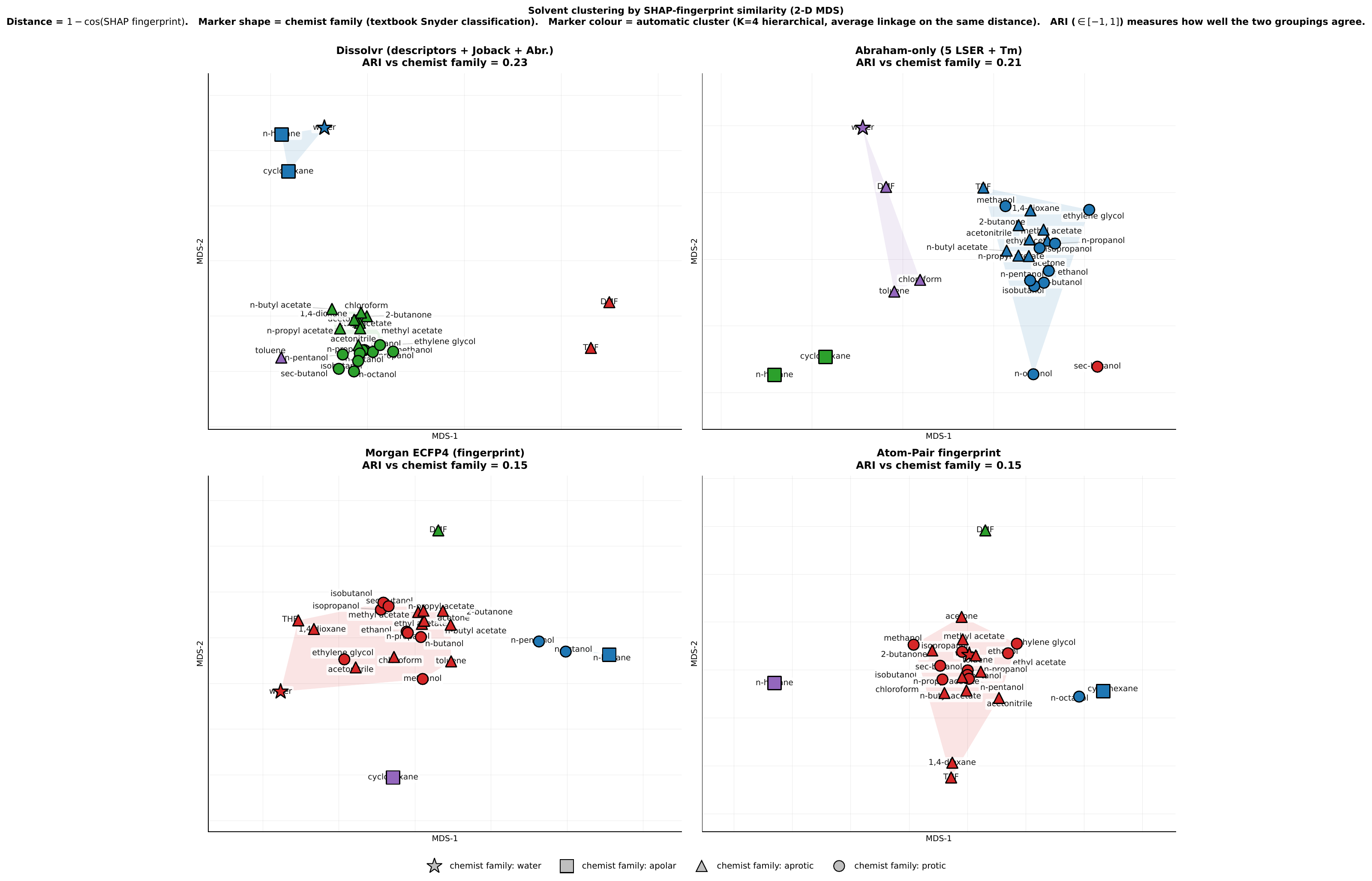}
\caption{\textbf{Solvent maps across four featurizers.}  Same MDS-on-SHAP
construction as Figure~\ref{fig:interp_solvent_map} but with one panel per
featurizer.  Top row (\textit{descriptor / Abraham}) cleanly separates water,
alkanes, the aprotic-aromatic group (DMF / THF / chloroform / toluene), and a
large alcohol-and-ester cluster.  Bottom row (\textit{circular fingerprints})
is structurally driven -- it isolates DMF and cyclohexane as singletons, places
n-hexane next to long-chain alcohols, and dumps everything else into one mass
-- a chemistry-blind grouping.  ARI scores quantify the agreement gap
(\numval{0.21}--\numval{0.23} for descriptors vs.\ \numval{0.15} for
fingerprints).}
\label{fig:interp_dendrograms}
\end{figure}

On the descriptor representations the model partitions the solvents into four
chemistry-readable groups: water, alkanes, an aprotic-aromatic cluster (DMF,
THF, chloroform, toluene), and one large band that contains the alcohols,
esters, ketones, ethers, acetonitrile, and dioxane.  The first three groups
match the chemist's families one-to-one; the fourth is a genuine model
artefact -- under the descriptor view those 16 solvents really do have nearly
identical SHAP fingerprints, because the model's attribution mass on solvent
features is small (\(\sim\)\numval{20}\,\%, §\ref{app:interp_blockwise}) and
the residual variance is dominated by the solute.

The fingerprint-based maps tell a strikingly different story.  Morgan ECFP4
isolates DMF, cyclohexane, and water as solitary outliers and groups n-hexane
with n-pentanol and n-octanol -- chemically incoherent, because those latter
two are protic alcohols of distinct functional class.  This is structural
similarity (long carbon chains share Morgan bits) masquerading as chemical
similarity.  The Atom-Pair fingerprint behaves the same way.  The Adjusted
Rand Index quantifies the gap: \numval{0.21}--\numval{0.23} for the descriptor
/ Abraham maps vs.\ \numval{0.15} for the fingerprints.

This is the mechanistic complement of the Representation ablation's RMSE story.
The Representation ablation showed circular fingerprints trail descriptors by
\(\approx\) \numval{0.15} \RMSE{} on \texttt{ood} and \texttt{sc3\_gold}; here
we see \emph{why} -- the fingerprint model has no internal representation under
which water behaves differently from a long-chain alcohol -- and so it cannot
specialise its prediction the way a descriptor model can.

\subsection{Abraham/LSER axes per solvent}
\label{app:interp_abraham}

\begin{figure}[t]
\centering
\includegraphics[width=\linewidth]{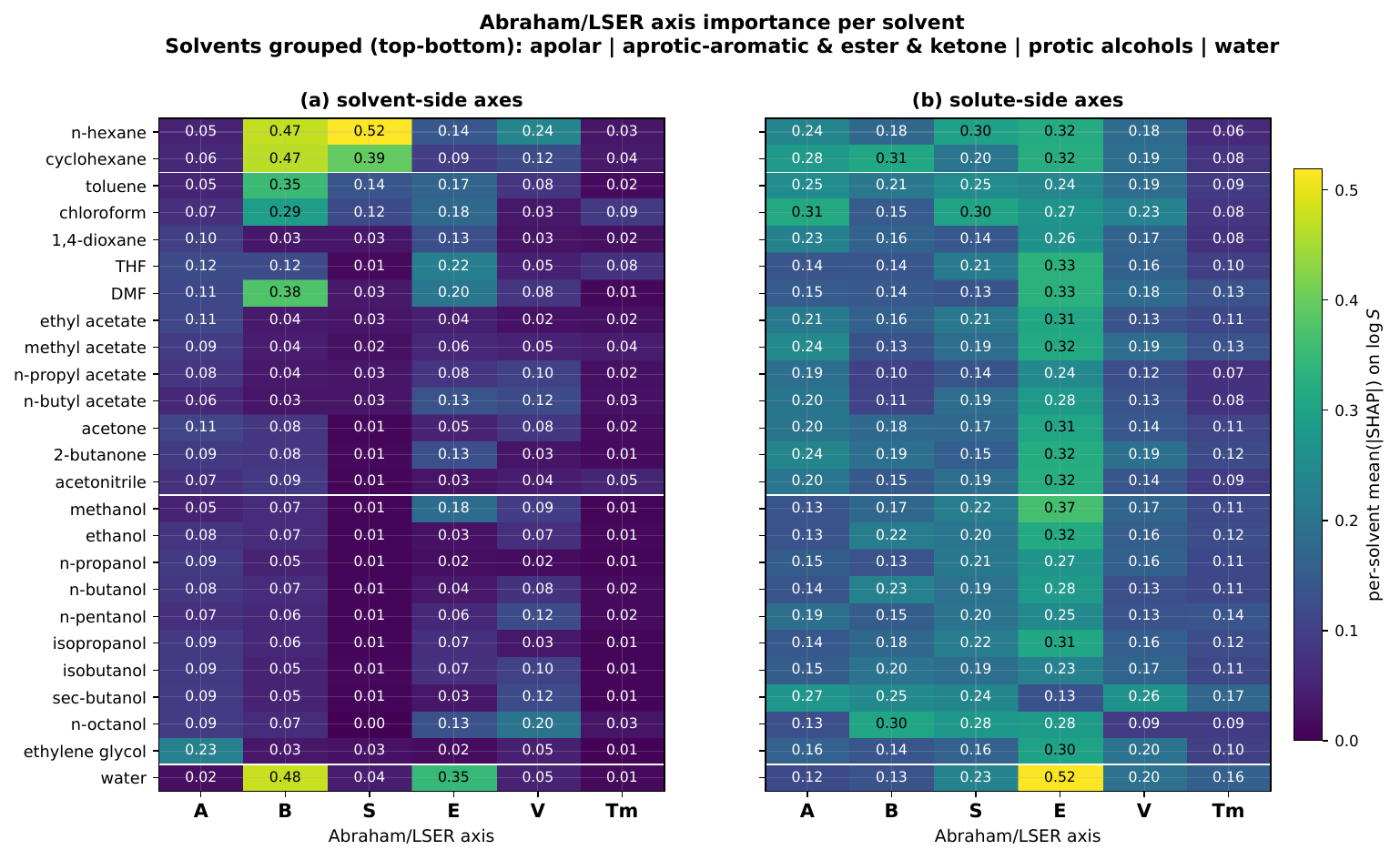}
\caption{\textbf{Abraham/LSER axis importance per solvent.}  Rows are solvents
grouped (top-bottom) into apolar, aprotic-aromatic \(+\)~ester~/~ketone,
protic alcohols, and water.  Columns are the six Abraham/LSER axes.  Cells are
mean(|SHAP|) for that (side, solvent, axis) cell on \texttt{eval}.  The panel
reads as a literal LSER table: alkanes have very large $B$ and $S$ on the
solvent side, water has a large solvent-side $E$ \emph{and} the largest
solute-side $E$ in the dataset, and DMF/toluene foreground solvent-side $B$.}
\label{fig:interp_abraham}
\end{figure}

The 6-feature Abraham-only representation makes the solvatochemistry unusually
transparent: every feature has a textbook physical meaning ($A$ = H-bond
acidity, $B$ = H-bond basicity, $S$ = dipolarity / polarisability, $E$ =
excess molar refractivity, $V$ = molar volume, $T_m$ = melting-point proxy).
Figure~\ref{fig:interp_abraham} shows the per-solvent breakdown of mean(|SHAP|)
along each axis, side-by-side for the solvent block (left) and the solute block
(right).

Several known-correct chemistry signals appear unprompted.  On the solvent
side, the \(B\) axis (H-bond basicity) is the dominant feature for n-hexane,
cyclohexane, water, DMF, and toluene -- exactly the solvents whose \(B\) value
is unusual.  On the solute side, the \(E\) axis (excess molar refractivity /
aromaticity proxy) dominates everywhere, peaking sharply at solvent~$=$~water
(\numval{0.52}) -- which matches the well-known dependence of aqueous
solubility on solute aromaticity.

\subsection{Pairwise interactions}
\label{app:interp_interactions}

\begin{figure}[t]
\centering
\includegraphics[width=\linewidth]{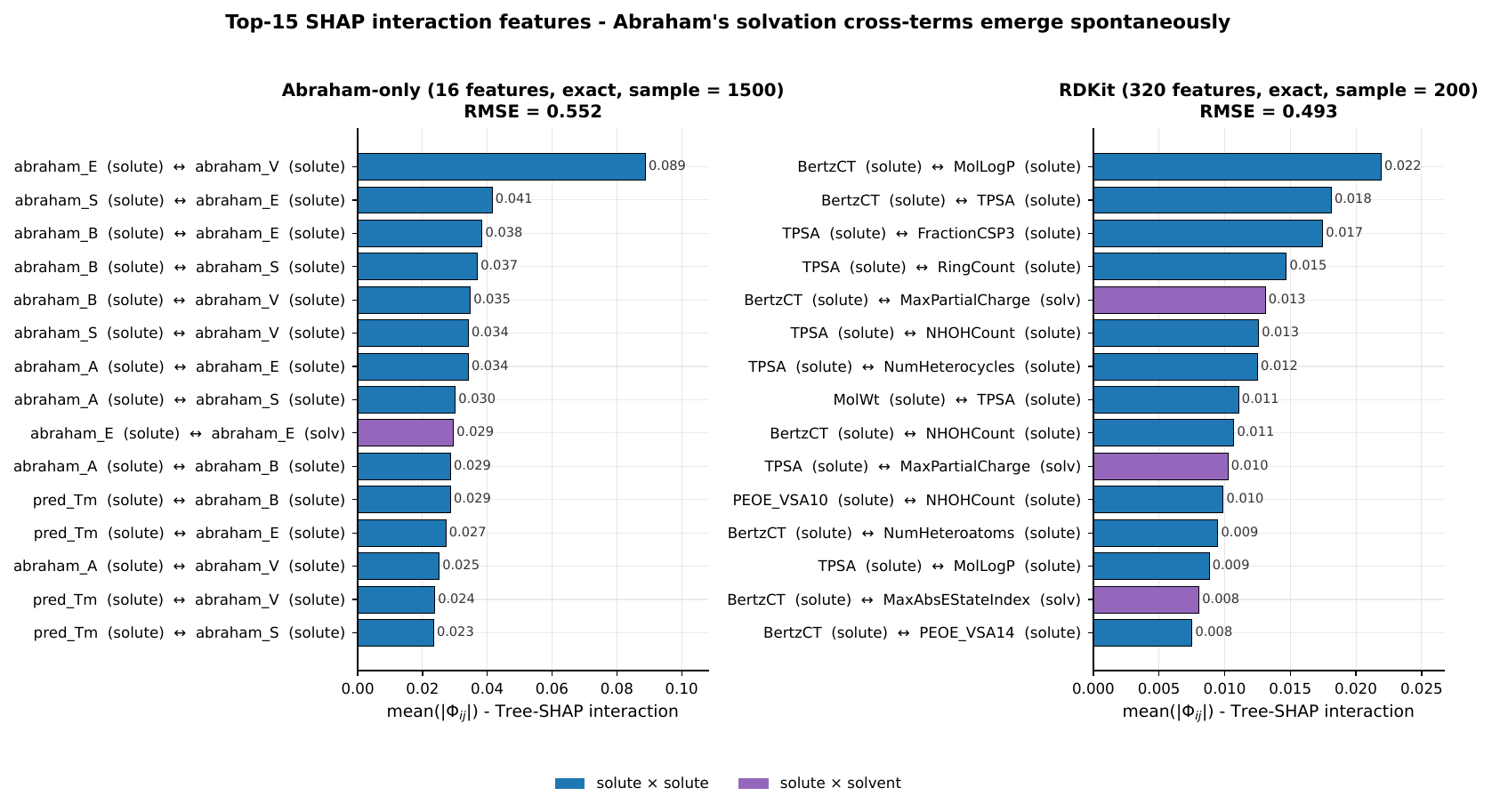}
\caption{\textbf{Top SHAP interaction features.}  Bars are
mean(|$\Phi_{ij}$|) on \texttt{eval}; colour encodes the block-pair
(solute$\times$solute, solute$\times$solvent, etc.).  On Abraham-only, the
top-8 pairs are exclusively within the solute LSER axes (textbook
within-molecule solvation coupling); cross-block solute$\times$solvent pairs
appear from rank 9 onward and are matched-axis ($E\!\times\!E$,
$A\!\times\!A$, $E\!\times\!B$) -- the off-diagonal LSER cross-terms.  On
RDKit, the top-4 pairs are within the GSE axes; the largest cross-block
interaction is \texttt{BertzCT(solute)}\(\times\)\texttt{MaxPartialCharge(solv)},
the discrete analogue of a partition coefficient term.}
\label{fig:interp_interactions}
\end{figure}

So far every analysis has been an additive (main-effect) decomposition of the
predictions.  Tree-SHAP also defines exact \emph{interaction} values $\Phi_{ij}$
that tell us how much of the prediction comes from the joint use of features $i$
and $j$, after subtracting the main effects.
Figure~\ref{fig:interp_interactions} shows the top-15 interaction pairs for
Abraham-only (left, exact, \numval{1500}-row \texttt{eval} sample) and RDKit
(right, exact, \numval{200}-row sample with
\texttt{tree\_limit}=\numval{150} to keep compute tractable).

The Abraham-only panel reproduces the structure of Abraham's solvation equation
almost exactly.  The top interaction
($E_{\mathrm{solute}}\!\times\!V_{\mathrm{solute}}$,
mean(|$\Phi$|) = \numval{0.089}) is the same coupling the LSER equation writes
as $v\,V + e\,E$.  Pairs 2--8 are all other within-LSER couplings
($S\!\times\!E$, $B\!\times\!E$, $B\!\times\!S$, etc.).  Cross-block
solute$\times$solvent matched-axis interactions appear from rank 9 onward
($E_{\mathrm{solute}}\!\times\!E_{\mathrm{solv}}$,
$A_{\mathrm{solute}}\!\times\!A_{\mathrm{solv}}$), which is exactly the
off-diagonal structure that LSER predicts.  No solvent-T or T-T interactions
appear in the top 20: temperature enters the model essentially additively.

The RDKit panel is the descriptor analogue.  The top four interactions are all
solute-side and span the GSE axes
(\texttt{BertzCT}\(\times\)\texttt{MolLogP},
\texttt{BertzCT}\(\times\)\texttt{TPSA},
\texttt{TPSA}\(\times\)\texttt{FractionCSP3},
\texttt{TPSA}\(\times\)\texttt{RingCount}).  The largest cross-block pair is
\texttt{BertzCT(solute)}\(\times\)\texttt{MaxPartialCharge(solv)} at rank 5:
the model gates the effect of solute complexity on solubility by how polar the
solvent is.

\subsection{Graph-based interpretation: GCN}
\label{app:interp_gcn}

The SHAP analysis above is unique to tree-based models.  For the dual-encoder
GCN we use occlusion attribution: for each (\textit{solute}, \textit{solvent},
\(T\)) triple in a sample of \numval{5000} \texttt{eval} rows, we compute the
model's prediction once with the full graph and once with each solute atom's
node features set to zero, recording the absolute change \(|\Delta\,\logS|\).
This gives \numval{85062} per-atom attribution scores, which we aggregate two
ways: by atom type (Figure~\ref{fig:row1_interp}a in the main paper) and by
BRICS fragment containing the atom (Figure~\ref{fig:row1_interp}b in the main
paper).

The atom-type ranking is led by sulfur and phosphorus: occluding a single
acyclic sulfur atom shifts the prediction by \(\approx\) \numval{1.13}
\logS{} on average -- about twice the effect of a typical carbon
(\numval{0.58}).  Sulfur and phosphorus appear relatively rarely in the dataset
(\numval{638} acyclic-S and \numval{104} acyclic-P out of
\(\sim\)\,\numval{85000} atom occurrences) but each occurrence carries large
weight, which matches the chemical intuition that sulfones, thiols, sulfonates,
phosphonates, etc.\ dramatically change solubility.  Aromatic carbon and
aliphatic ring-carbon have nearly identical mean \(|\Delta|\)
\(\approx\) \numval{0.58}, consistent with the GCN being built on seven plain
atom features without an explicit \(\pi\)-system descriptor: aromaticity is
encoded only as a topology cue.

The BRICS fragment ranking is the chemically actionable view.  Every fragment
in the top fifteen is one a medicinal chemist would immediately list as a
solubility-relevant motif: aromatic core (benzene, phenol, naphthalene,
chlorobenzene, pyridine, nitrobenzene), H-bonding groups (carboxylic acid,
amide-like, phenolic OH), small carbonyls, alcohol and amine motifs, and a
piperazine ring.  No uninterpretable fragment appears in the top twenty, which
is a strong indication that the GCN is using message-passing to aggregate
chemistry-meaningful subgraphs even though it was never explicitly told what
they are.

\begin{figure}[t]
\centering
\includegraphics[width=\linewidth]{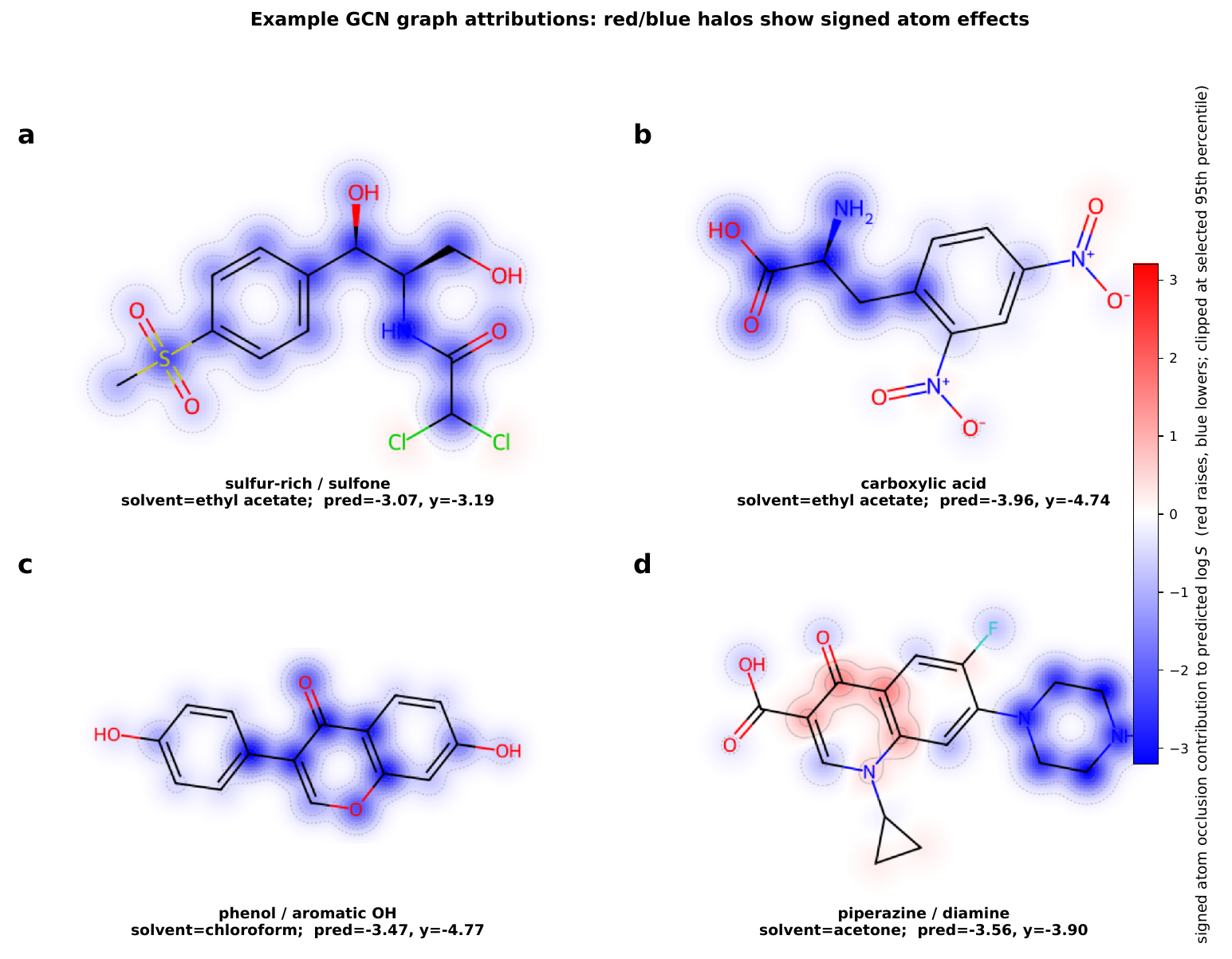}
\caption{\textbf{Example GCN graph attributions.}  Four solutes chosen from the
\texttt{eval} split to expose the motifs that dominate the aggregate BRICS
ranking: sulfur/sulfone, carboxylic acid / nitro-aromatic, phenol/aromatic OH,
and piperazine/diamine.  The underlying attribution is signed atom occlusion:
red regions increase the predicted \logS{}; blue regions decrease it.  The maps
are qualitative but coherent: acidic, nitro, sulfone, and amide-like atoms are
high-leverage; aromatic scaffolds often provide broad negative fields; and
heteroatom-rich fragments such as piperazine or carboxylate dominate local
prediction changes.}
\label{fig:interp_gcn_examples}
\end{figure}

The aggregate tables make the result statistically stable, but they hide what a
graph explanation looks like on a single molecule.  We therefore draw four
representative solutes in Figure~\ref{fig:interp_gcn_examples}.  For each
molecule we run the same occlusion experiment atom-by-atom, but keep the sign:
\(\Delta_i = f(G) - f(G \setminus i)\), where \(G\setminus i\) means zeroing
atom \(i\)'s node features while leaving the topology intact.

These molecule-level maps also clarify what the aggregate BRICS plot cannot.
The same functional group can participate in different directions depending on
its scaffold and solvent: a carboxylate can be blue in a nitroaromatic acid
because it lowers the prediction relative to the rest of that molecule, while
the same carboxylate fragment still appears globally important in
Figure~\ref{fig:row1_interp} because its occlusion magnitude is large.  This is
why we report both views: signed maps are mechanistic examples; aggregate BRICS
counts are the stable population-level result.

\subsection{Caveats and reproducibility}
\label{app:interp_caveats}

\paragraph{Caveats.}  All interpretations above are computed at
seed\,$=$\,\numval{42}.  The Representation ablation shows seed-to-seed \RMSE{}
jitter is \(\sim\)\numval{0.005}; we therefore expect the top-3 features per
featurizer to be stable across seeds, but ranks beyond \(\sim\)\numval{10}
could reshuffle.  Tree-SHAP interaction values are computed in full only for
Abraham-only (16 features, exact); RDKit uses a smaller \numval{200}-row sample
with \texttt{tree\_limit}\,$=$\,\numval{150}, and Mordred / Morgan / Atom-Pair
/ MACCS interaction values are skipped because the $O(N \cdot T \cdot F^2)$
cost is prohibitive on a single CPU.  Signed SHAP directions are
rank-correlations between feature values and SHAP values, so they should be
read as fitted-model associations, not causal derivatives.  GCN attribution
uses ``soft'' node-feature occlusion (zero the node features, keep the
topology), which preserves message-passing well-definedness.

\paragraph{Reproducibility.}  All scripts, intermediate SHAP arrays
($\sim$\numval{420}\,MB), 85 figures, and per-solvent CSV summaries live in the released repository.

%% file: sections/appendix_data_scaling.tex
\section{Data scaling: experimental protocol and extended results}
\label{app:data_scaling}

This appendix contains the full experimental protocol, mathematical
definitions, and supporting figures for the data-scaling study summarised
in §\ref{sec:data_scaling} of the main paper.

\subsection{Split-specific aleatoric RMSE floors}
\label{app:scaling_floors}

For each benchmark split $s$ (Train, Eval, OOD, Gold-tier analogue
\texttt{sc3\_gold}), we estimate an RMSE-comparable aleatoric floor from
duplicate $(\text{solute},\text{solvent},\mathrm{round}(T,0))$ triples:
observations within the same triple $g$ are treated as repeated draws of
the same underlying label.
Let $\bar{y}_g$ be the mean label in group $g$, $n_g$ the number of
observations in $g$, and $N=\sum_g n_g$ the total count of observations
participating in duplicate groups on split $s$.
We define
\begin{equation}
  \bigl(\varepsilon_A^{(s)}\bigr)^2
  \;=\; \frac{1}{N}\sum_{g}\sum_{i\in g}\bigl(y_i-\bar{y}_g\bigr)^2 .
  \label{eq:eps_split}
\end{equation}
Values are reported in Table~\ref{tab:scaling_floors}; they serve as
split-specific references when interpreting the asymptotes of fitted
scaling curves below.

\input{tables/scaling_floors}

\subsection{Empirical scaling curves}
\label{app:scaling_curves}

\begin{figure}[t]
  \centering
  \includegraphics[width=\linewidth]{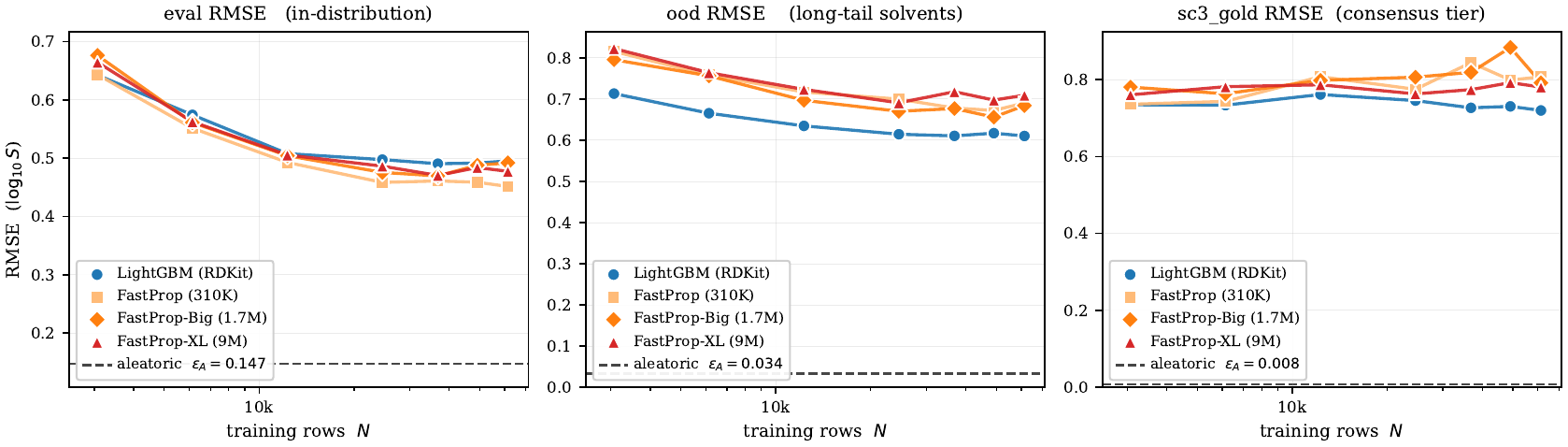}
  \caption{Test RMSE vs.\ training-set fraction for representative models
    (seven fractions $\times$ fixed seeds).  All models plateau well
    before the full training set, indicating that performance is limited
    by representation capacity rather than data volume.}
  \label{fig:data_scaling_curves}
\end{figure}

\begin{figure}[t]
  \centering
  \includegraphics[width=\linewidth]{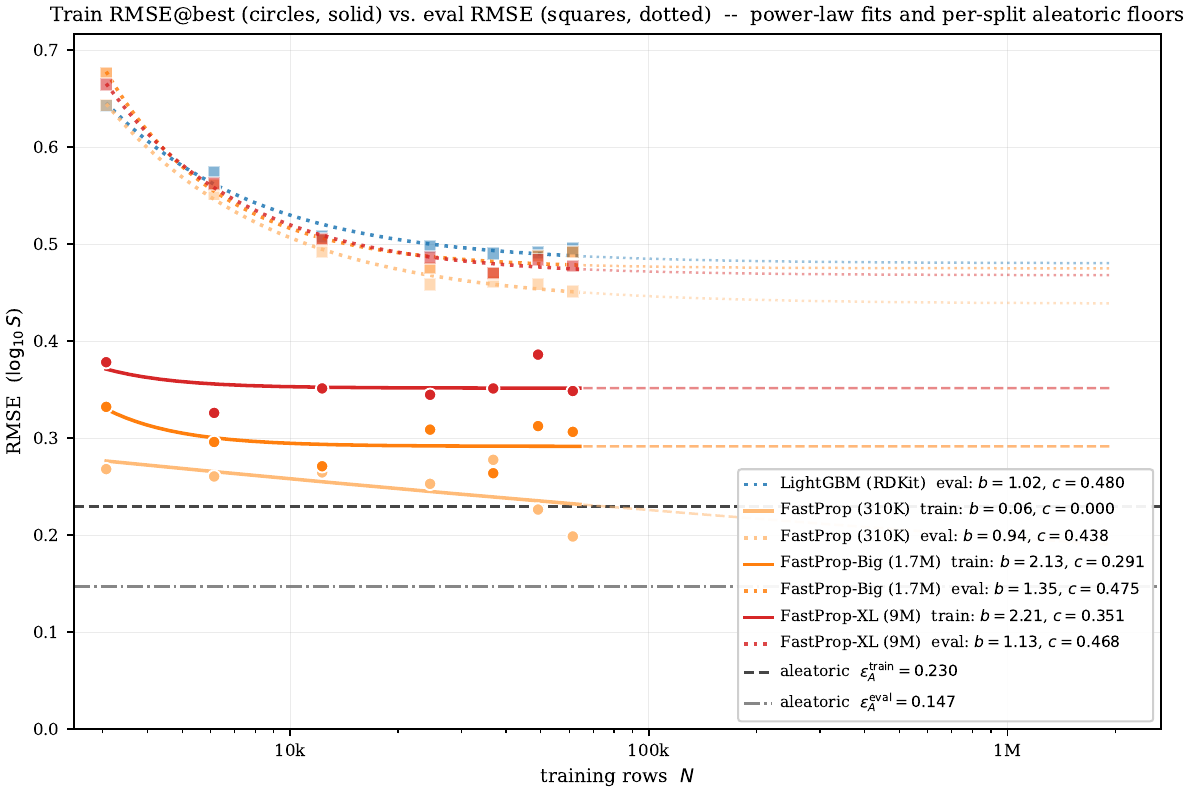}
  \caption{Train vs.\ test RMSE across fractions (diagnostic for
    overfitting vs.\ noise floor).  The train--test gap stabilises at
    large $N$, ruling out overfitting as the primary driver of the
    test-RMSE plateau.}
  \label{fig:data_scaling_train_vs_val}
\end{figure}

\begin{figure}[t]
  \centering
  \includegraphics[width=\linewidth]{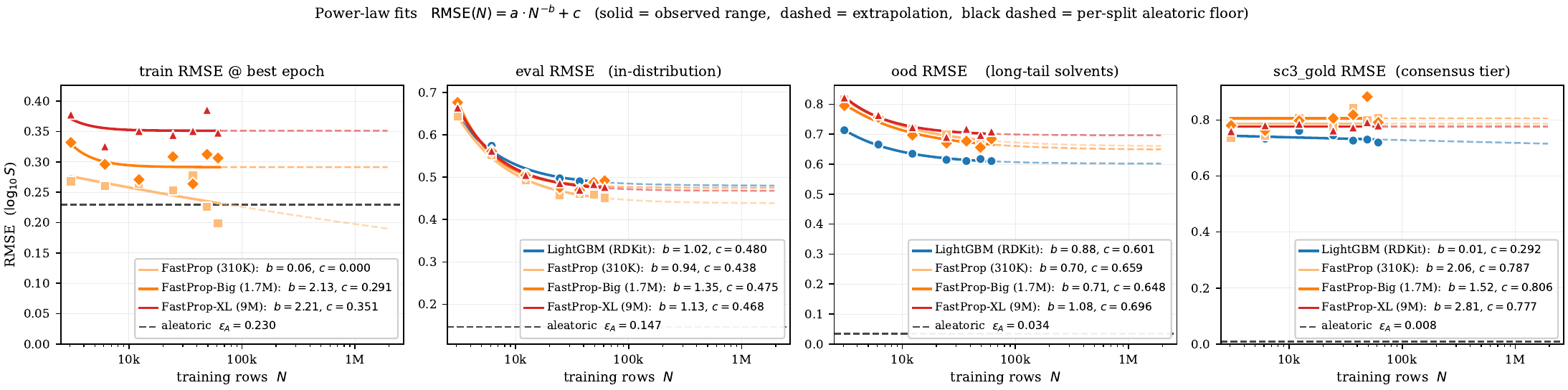}
  \caption{Saturating power-law fits overlaid on empirical scaling curves.
    Dashed horizontal lines mark the split-specific aleatoric floors
    $\varepsilon_A^{(s)}$ (Table~\ref{tab:scaling_floors}).
    Every fitted asymptote $c$ (dotted lines) lies strictly above its
    floor, confirming that no finite training set closes the gap within
    this representation class (Table~\ref{tab:scaling_extrapolation}).}
  \label{fig:data_scaling_lawfits}
\end{figure}

We subsample the training pool at fractions
$\{0.05,0.10,0.20,0.40,0.60,0.80,1.00\}$ (stratified by solvent) and
retrain representative models---\textsc{LightGBM} on RDKit descriptors
and three \textsc{FastProp} MLP widths ($310$K, $3.0$M, and $9.0$M
parameters)---holding all evaluation splits fixed across replicates.
Figure~\ref{fig:data_scaling_curves} shows test RMSE vs.\ training-set
size; Figure~\ref{fig:data_scaling_train_vs_val} decomposes train vs.\
test behaviour as a diagnostic for overfitting; and
Figure~\ref{fig:data_scaling_lawfits} overlays the saturating power-law
fits together with the per-split aleatoric floors.

\subsection{Power-law asymptotes}
\label{app:scaling_powerlaw}

Table~\ref{tab:scaling_fits} lists least-squares fits of
$\mathrm{RMSE}(N)=a N^{-b}+c$ on the seven-point scaling curves.
The intercept $c$ is interpreted as the best RMSE achievable at infinite
training data \emph{for the fixed representation and architecture}; it
should be compared to $\varepsilon_A^{(s)}$ from Eq.~\ref{eq:eps_split},
not to the global inter-laboratory $\varepsilon_A$.

\input{tables/scaling_fits}

\subsection{Extrapolated crossing with the aleatoric floor}
\label{app:scaling_q4}

Given a fitted asymptote $c$, split floor $\varepsilon_A^{(s)}$, and
tolerance $\delta=0.05$~log~S, the hypothetical training-set size at which
the power-law model would match the aleatoric reference satisfies
\begin{equation}
  a\,(N^\star_{\varepsilon_A})^{-b} + c
  \;\le\; \varepsilon_A^{(s)} + \delta .
  \label{eq:nstar}
\end{equation}
Whenever $c>\varepsilon_A^{(s)}+\delta$, no finite $N^\star_{\varepsilon_A}$
exists within the saturating model---Table~\ref{tab:scaling_extrapolation}
marks this case as $\infty$.
We additionally report $N^\star_{95}$, the size at which RMSE reaches
within $5\%$ of the model's \emph{own} asymptote $c$, as a data-efficiency
summary independent of $\varepsilon_A^{(s)}$.

\input{tables/scaling_extrapolation}

\paragraph{Relation to §\ref{sec:ablations}.}
The headline conclusion cited there carries directly from
Table~\ref{tab:scaling_extrapolation}: within this representation class,
fitted asymptotes lie comfortably above the duplicate-triple floors on
Eval and OOD, so---under the power-law model---error cannot be trained
through the measurement-limited regime without changing features or
architecture.

%% file: tables/scaling_floors.tex
\begin{table}[h]
\centering\small
\caption{Per-split aleatoric RMSE floors $\epsA$ from
  Eq.~\ref{eq:eps_split}, computed on duplicate
  $(\text{solute}, \text{solvent}, \mathrm{round}(T,0))$ triples within
  each split.  $n_{\text{trip}}$ is the number of duplicate triples;
  $n_{\text{obs}}$ is the total number of observations participating
  in those triples (each contributes one residual against its group
  mean).  Median per-triple std and the $90$th percentile capture the
  shape of the within-triple disagreement distribution: medians are
  sub-millimolar everywhere, and the heavy right tail is what drives
  $\epsA$ above the median.  $\epsA$ is RMSE-comparable to the model
  RMSE reported on the same split.}
\label{tab:scaling_floors}
\begin{tabular}{@{}lrrrrrr@{}}
\toprule
split & $n_{\text{trip}}$ & $n_{\text{obs}}$ & $\epsA$ (RMSE) & MAE floor & $P_{50}$ & $P_{90}$ \\
\midrule
\texttt{train}    & \numval{905} & \numval{1\,839} & \numval{0.230} & \numval{0.074} & \numval{0.005} & \numval{0.430} \\
\texttt{eval}     & \numval{101} & \numval{202}    & \numval{0.147} & \numval{0.038} & \numval{0.005} & \numval{0.026} \\
\texttt{ood}      & \numval{83}  & \numval{166}    & \numval{0.034} & \numval{0.009} & \numval{0.002} & \numval{0.024} \\
\texttt{sc3\_gold} & \numval{490} & \numval{988}   & \numval{0.008} & \numval{0.004} & \numval{0.002} & \numval{0.012} \\
\bottomrule
\end{tabular}
\end{table}

%% file: tables/scaling_fits.tex
\begin{table}[h]
\centering\small
\caption{Power-law fits $\mathrm{RMSE}(N) = a \cdot N^{-b} + c$ per
  (model, evaluation split), plus the train-side fit (only available for
  the deep models; LightGBM does not save a per-fraction train-RMSE
  diagnostic).  $b$ is the scaling exponent (larger = more bang per
  data row).  $c$ is the model's irreducible asymptote (smaller =
  better in the limit of infinite data).  $R^{2}$ is the coefficient
  of determination of the fit on the seven $(N, \mathrm{RMSE})$ points.
  The asymptote $c$ is the column to read for Q4: it is the model's
  best achievable RMSE on this split, given infinite data and the same
  representation.}
\label{tab:scaling_fits}
\begin{tabular}{@{}llrrrr@{}}
\toprule
model              & split           & $a$           & $b$    & $c$            & $R^{2}$ \\
\midrule
\textsc{LightGBM}  & \texttt{eval}    & \numval{588.43} & \numval{1.02} & \numval{0.480} & \numval{0.98} \\
\textsc{LightGBM}  & \texttt{ood}     & \numval{131.38} & \numval{0.88} & \numval{0.601} & \numval{0.99} \\
\textsc{LightGBM}  & \texttt{sc3\_gold}& \numval{0.49}   & \numval{0.01} & \numval{0.292} & \numval{0.13} \\
\midrule
\textsc{FastProp}  & \texttt{eval}    & \numval{382.15} & \numval{0.94} & \numval{0.438} & \numval{0.99} \\
\textsc{FastProp}  & \texttt{ood}     & \numval{45.07}  & \numval{0.70} & \numval{0.659} & \numval{0.98} \\
\textsc{FastProp}  & \texttt{sc3\_gold}& \numval{$\sim$0}& \numval{2.06} & \numval{0.787} & --- \\
\textsc{FastProp}  & \texttt{train}   & \numval{0.44}   & \numval{0.06} & \numval{$\sim 0$} & \numval{0.36} \\
\midrule
\textsc{FastProp-Big}& \texttt{eval}  & \numval{10\,262.47}& \numval{1.35}& \numval{0.475} & \numval{0.98} \\
\textsc{FastProp-Big}& \texttt{ood}   & \numval{47.26}  & \numval{0.71} & \numval{0.648} & \numval{0.94} \\
\textsc{FastProp-Big}& \texttt{sc3\_gold}& \numval{$\sim$0}& \numval{1.52}& \numval{0.806} & --- \\
\textsc{FastProp-Big}& \texttt{train} & \numval{$10^{6}$}& \numval{2.13}& \numval{0.291} & \numval{0.35} \\
\midrule
\textsc{FastProp-XL} & \texttt{eval}  & \numval{1\,740.30}& \numval{1.13}& \numval{0.468} & \numval{0.99} \\
\textsc{FastProp-XL} & \texttt{ood}   & \numval{734.60} & \numval{1.08} & \numval{0.696} & \numval{0.95} \\
\textsc{FastProp-XL} & \texttt{sc3\_gold}& \numval{0.83}& \numval{2.81} & \numval{0.777} & --- \\
\textsc{FastProp-XL} & \texttt{train} & \numval{$10^{6}$}& \numval{2.21}& \numval{0.351} & \numval{0.13} \\
\bottomrule
\end{tabular}
\end{table}

%% file: tables/scaling_extrapolation.tex
\begin{table}[h]
\centering\small
\caption{Q4 extrapolation table.  For every (model, evaluation split)
  pair we report the asymptote $c$ from
  Table~\ref{tab:scaling_fits}, the per-split aleatoric floor $\epsA$
  from Table~\ref{tab:scaling_floors}, the gap $\Delta = c - \epsA$,
  the training-set size $N^{\star}_{95}$ at which RMSE comes within
  $5\%$ of the model's own asymptote (a moderate, achievable target),
  and the size $N^{\star}_{\epsA}$ that would bring RMSE to within
  $0.05$ logS of the aleatoric floor (Eq.~\ref{eq:nstar}).
  Entries marked $\infty$ have $c \geq \epsA + 0.05$, so no finite $N$
  meets the target within the saturating power-law model.  All
  $N^{\star}_{\epsA}$ are $\infty$ -- the central Q4 finding.}
\label{tab:scaling_extrapolation}
\begin{tabular}{@{}llrrrrr@{}}
\toprule
model              & split             & $c$ (asymptote) & $\epsA$ & $\Delta = c - \epsA$ & $N^{\star}_{95}$ & $N^{\star}_{\epsA}$ \\
\midrule
\textsc{LightGBM}  & \texttt{eval}     & \numval{0.480} & \numval{0.147} & \numval{0.333} & \numval{59\,000}  & $\infty$ \\
\textsc{LightGBM}  & \texttt{ood}      & \numval{0.601} & \numval{0.034} & \numval{0.567} & \numval{93\,000}  & $\infty$ \\
\textsc{LightGBM}  & \texttt{sc3\_gold}& \numval{0.292} & \numval{0.008} & \numval{0.283} & --- (flat fit)    & $\infty$ \\
\midrule
\textsc{FastProp}  & \texttt{eval}     & \numval{0.438} & \numval{0.147} & \numval{0.292} & \numval{76\,000}  & $\infty$ \\
\textsc{FastProp}  & \texttt{ood}      & \numval{0.659} & \numval{0.034} & \numval{0.625} & \numval{217\,000} & $\infty$ \\
\textsc{FastProp}  & \texttt{sc3\_gold}& \numval{0.787} & \numval{0.008} & \numval{0.779} & --- (flat fit)    & $\infty$ \\
\midrule
\textsc{FastProp-Big}& \texttt{eval}   & \numval{0.475} & \numval{0.147} & \numval{0.328} & \numval{28\,000}  & $\infty$ \\
\textsc{FastProp-Big}& \texttt{ood}    & \numval{0.648} & \numval{0.034} & \numval{0.614} & \numval{213\,000} & $\infty$ \\
\textsc{FastProp-Big}& \texttt{sc3\_gold}& \numval{0.806}& \numval{0.008}& \numval{0.798} & --- (flat fit)    & $\infty$ \\
\midrule
\textsc{FastProp-XL} & \texttt{eval}   & \numval{0.468} & \numval{0.147} & \numval{0.321} & \numval{44\,000}  & $\infty$ \\
\textsc{FastProp-XL} & \texttt{ood}    & \numval{0.696} & \numval{0.034} & \numval{0.662} & \numval{50\,000}  & $\infty$ \\
\textsc{FastProp-XL} & \texttt{sc3\_gold}& \numval{0.777}& \numval{0.008}& \numval{0.769} & --- (flat fit)    & $\infty$ \\
\bottomrule
\end{tabular}
\end{table}

%% file: sections/appendix_transfer.tex
\section{Transfer learning: extended material}
\label{app:transfer}

The data-scaling analysis of §\ref{sec:data_scaling} established that
extra \scthree{} solubility data does not, on its own, close the model
gap to the aleatoric floor.  A complementary lever is to share
inductive bias from a different but \emph{adjacent} chemical task that
has more data than \scthree{} can provide.  We test the strongest
candidate: the \textsc{CombiSolv-QM} dataset of
\citet{vermeire2021transfer}, $\sim$$10^6$ \textsc{cosmo-rs} solvation
free energies $\Delta G_{\text{solv}}$ at $T = \numval{298.15}$~K
spanning $11{,}029$ solutes and $284$ solvents.  CombiSolv-QM has the
exact $({\text{solute}}, {\text{solvent}})$ pair structure as
\scthree{}, two orders of magnitude more rows than \scthree{}-train,
and is purely computational so its labels carry essentially no
experimental noise.  If pretraining a regressor on
\textsc{CombiSolv-QM} and fine-tuning on a fraction of \scthree{}-train
beats training on the same fraction from scratch, we have evidence
that the chemistry signal in $\Delta G_{\text{solv}}$ transfers to
\logS{}.

We split this question into two sub-questions to control for the one
obvious confound between the two tasks: \scthree{} spans
$T \in [\numval{243}, \numval{383}]$~K, whereas
\textsc{CombiSolv-QM} is at a single temperature.

\begin{itemize}[leftmargin=*]
  \item §\ref{app:transfer_multiT} -- the natural setup: pretrain at
        $\numval{298.15}$~K, fine-tune on the full multi-temperature
        \scthree{}-train.  This measures the \emph{practical} value
        of \textsc{CombiSolv-QM} pretraining for our benchmark.
  \item §\ref{app:transfer_298K} -- the temperature-isolated setup:
        for every $({\text{solute}}, {\text{solvent}})$ pair in
        \texttt{interim/04\_fits.csv} (the per-pair Apelblat /
        Van't~Hoff fits used in the data-cleaning pipeline) we
        evaluate the fit at exactly $T = \numval{298.15}$~K, giving
        a single-temperature variant of \scthree{} where pretraining
        and fine-tuning live on the same temperature axis.  This
        isolates the chemistry signal from temperature dependence.
\end{itemize}

Both setups use a single, identical FastProp~\citep{burns2025fastprop}
architecture and an identical fine-tuning recipe; the only knob that
changes between scratch and qm protocols is whether the trunk weights
are random-initialised or loaded from the
\textsc{CombiSolv-QM}-pretrained checkpoint.

\subsection{Experimental design}
\label{app:transfer_setup}

\paragraph{Architecture.}  The model is the same \textsc{FastProp} MLP
that appears in the headline benchmark of §\ref{sec:baselines}: three
$[\text{Linear} \to \text{BatchNorm} \to \text{ReLU} \to
\text{Dropout}(0.1)]$ blocks of widths $(\numval{512}, \numval{256},
\numval{128})$ followed by a single $\text{Linear}(\numval{128}, 1)$
regression head.  The hidden width and dropout are taken verbatim
from \texttt{configs/best\_hps.json[\textsf{fastprop}]}.  Total
parameters: $\numval{330497}$.  Identity scaling --- the trunk has no
output normalisation --- so the transfer protocol can swap the head
to predict either $\Delta G_{\text{solv}}$ (kcal\,mol$^{-1}$) or
\logS{} without re-balancing the loss.

\paragraph{Featurisation.}  Both tasks use the SC$^3$ benchmark's
RDKit-2D pipeline (§\ref{sec:baselines}): \numval{158} 2D descriptors
per molecule, concatenated for solute and solvent (\numval{316}
features), plus four temperature features $T/300$, $1000/T$,
$(T/300)^2$, $\ln(T/300)$ ($\numval{320}$ features total).  For
\textsc{CombiSolv-QM} every row is at $T = \numval{298.15}$~K, so the
four temperature features are constant during pretraining.

\paragraph{Pretraining data.}  We use the cleaned
\textsc{CombiSolv-QM} release of
\citet{vermeire2021transfer}.  The original $\numval{999743}$ rows
are pair-level filtered against the \scthree{} holdouts
(\texttt{bench\_eval}, \texttt{bench\_ood}, \texttt{sc3/gold},
\texttt{sc3/silver}, \texttt{sc3/bronze}): we drop any row whose
\emph{canonical} $({\text{solute SMILES}}, {\text{solvent SMILES}})$
pair appears in any holdout.  Single-side overlap (same solute in a
different solvent, or vice versa) is \emph{not} counted as leakage,
following \citet{vermeire2021transfer}: knowing
$\Delta G_{\text{solv}}(\text{aspirin, ethanol})$ does not tell the
model the solubility of aspirin in hexane.  This drops only
$\numval{113}$ rows ($0.011$\%); the cleaned
\textsc{CombiSolv-QM} has $\numval{999630}$ rows.  Canonicalisation
uses RDKit \texttt{MolToSmiles} with
\texttt{canonical=True, isomericSmiles=False}.  The leakage report
is preserved at \texttt{Ablations/Transfer/data/leakage\_report.md}.

\paragraph{Pretraining recipe.}  The trunk is trained for up to
$\numval{60}$ epochs (in practice all five seeds early-stop) with
batch size $\numval{1024}$, Adam at $\text{lr}=\numval{5e-4}$,
weight decay $\numval{1e-5}$, plateau-based learning-rate decay
(factor $0.5$, patience $\numval{3}$), and early stopping with
patience $\numval{6}$ on a $5$\% held-out validation slice
($\numval{49982}$ rows).  Final pretrain validation
\RMSE{} on $\Delta G_{\text{solv}}$ is
$\numval{0.250}$--$\numval{0.258}$~kcal\,mol$^{-1}$ across the five
seeds, well below the chemical-accuracy threshold of
$\numval{1}$~kcal\,mol$^{-1}$.

\paragraph{Fine-tuning recipe.}  For each
$(\text{protocol}, \text{variant}, \text{fraction}, \text{seed})$
cell we initialise a FastProp model, optionally load the pretrained
trunk weights, replace the regression head with a fresh
$\text{Linear}(128, 1)$, and fine-tune.  The hyperparameters mirror
the \texttt{fastprop} config from
\texttt{configs/best\_hps.json}: Adam at lr $\numval{5e-4}$,
batch~$\numval{256}$, max~$\numval{300}$ epochs, patience
$\numval{40}$, plateau lr-decay (patience $\numval{15}$).  We use
\textsc{full} fine-tuning (every parameter trainable) and a
\textsc{head\_only} variant that freezes the trunk and trains only
the $\numval{129}$-parameter linear head.  Fractions are subsampled
\emph{stratified by solvent name} so even the $5$\% slice covers
every common solvent.

\paragraph{Input normalisation.}  Both protocols standardise input
features as $X' = (X - \mu) / \sigma$.  In the \textsc{qm} protocol,
$(\mu, \sigma)$ come from the pretraining set so the trunk sees
inputs in the scale it was trained on.  In the \textsc{scratch}
protocol, $(\mu, \sigma)$ come from the \emph{full} \scthree{}-train
set, not the subsample.  This second choice is essential: at small
fractions some RDKit columns happen to be all-zero in the random
sub\-sample but non-zero on the held-out splits, so standardising
with the subsample's near-zero $\sigma$ produces values
$\sim \! 10^{8}$ on the OOD set, which blow up the first BatchNorm
forward pass and trap early-stopping at epoch $1$.  We also patch
the normalisation to substitute identity scaling
$(\mu \! = \! 0, \sigma \! = \! 1)$ for any column whose
$\sigma < \numval{1e-3}$ in the reference set; this is a no-op for
columns that vary, and prevents the explosion otherwise.  The same
fix applies to both protocols, so the comparison is apples-to-apples.

\paragraph{BatchNorm calibration.}  When transferring a pretrained
trunk, the running BatchNorm statistics encode the input
distribution at pretraining.  Before the first eval pass we run one
no-grad forward pass over the fine-tune training data in
\texttt{train()} mode to refresh the running stats; otherwise the
stale running mean corrupts the eval-mode val-loss signal that
drives early stopping.  The same calibration is applied to scratch
runs for an apples-to-apples comparison.

\paragraph{Grid.}  Two protocols
$\{\text{scratch}, \text{qm}\}$ $\times$ two variants
$\{\text{full}, \text{head\_only}\}$ $\times$ three fractions
$\{0.05, 0.25, 1.0\}$ $\times$ five seeds
$\{42, 101, 123, 456, 789\}$ $= 60$ fine-tune runs per dataset.  We
report results on three held-out splits of \scthree{}: \texttt{eval}
(in-distribution, the same $\numval{25}$ solvents as
\scthree{}-train), \texttt{ood} (long-tail solvents not in the
training set, $\numval{146}$ unique solvents), and the
\texttt{sc3\_gold} consensus tier (rows with $\geq 2$ measurements
that agree to $\le 0.1$ \logSunit{}; the cleanest test set in the
benchmark).

\paragraph{Sanity check.}  Reproducing the headline FastProp
baseline at $100$\% \scthree{}-train and seed $42$ inside this
codebase gives \texttt{eval} \RMSE{} $\numval{0.4620 \pm 0.0036}$
(5 seeds, scratch/full), within $\numval{0.003}$~\logS{} of the
$\numval{0.4645}$ value reported in the main benchmark
(§\ref{sec:baselines}).  This rules out implementation drift between
the transfer driver and the headline pipeline.

\subsection{Multi-temperature setup}
\label{app:transfer_multiT}

We pretrain at $T \! = \! \numval{298.15}$~K and fine-tune on the
full multi-temperature \scthree{}-train ($\numval{61403}$ rows
spanning $\numval{243}$--$\numval{383}$~K).  This is the practical
setting: the pretrained representation does not directly know about
temperature dependence, so it has to be learned from \scthree{} on
top of whatever chemistry signal the trunk inherits.

Table~\ref{tab:transfer_multiT} reports the headline numbers.

\begin{table}[h]
\centering
\small
\caption{\textbf{Multi-T transfer: \RMSE{} (mean $\pm$ std over 5
seeds) for the \textsc{full}-fine-tune protocol.}  \scthree{}-train
spans \numval{243}--\numval{383}~K; \textsc{CombiSolv-QM}
pretraining is at \numval{298.15}~K only.  Lower is better; bold
marks the per-cell winner.  Negative $\Delta$
($=\text{qm}-\text{scratch}$) means \textsc{qm} pretraining helps.}
\label{tab:transfer_multiT}
\begin{tabular}{l l c c c}
\toprule
Split & Fraction & scratch & qm & $\Delta$ \\
\midrule
\multirow{3}{*}{\texttt{eval}}
  & $5$\%   & $0.661 \pm 0.025$ & $\mathbf{0.616 \pm 0.010}$ & $-0.045$ \\
  & $25$\%  & $0.484 \pm 0.013$ & $\mathbf{0.476 \pm 0.008}$ & $-0.008$ \\
  & $100$\% & $0.462 \pm 0.004$ & $\mathbf{0.459 \pm 0.008}$ & $-0.003$ \\
\midrule
\multirow{3}{*}{\texttt{ood}}
  & $5$\%   & $0.812 \pm 0.022$ & $\mathbf{0.751 \pm 0.022}$ & $-0.061$ \\
  & $25$\%  & $0.683 \pm 0.010$ & $\mathbf{0.650 \pm 0.018}$ & $-0.033$ \\
  & $100$\% & $0.672 \pm 0.012$ & $\mathbf{0.653 \pm 0.009}$ & $-0.019$ \\
\midrule
\multirow{3}{*}{\texttt{sc3\_gold}}
  & $5$\%   & $0.890 \pm 0.106$ & $\mathbf{0.784 \pm 0.050}$ & $-0.106$ \\
  & $25$\%  & $0.884 \pm 0.107$ & $\mathbf{0.784 \pm 0.032}$ & $-0.100$ \\
  & $100$\% & $0.805 \pm 0.036$ & $\mathbf{0.755 \pm 0.007}$ & $-0.050$ \\
\bottomrule
\end{tabular}
\end{table}

\begin{figure}[h]
\centering
\includegraphics[width=0.96\linewidth]{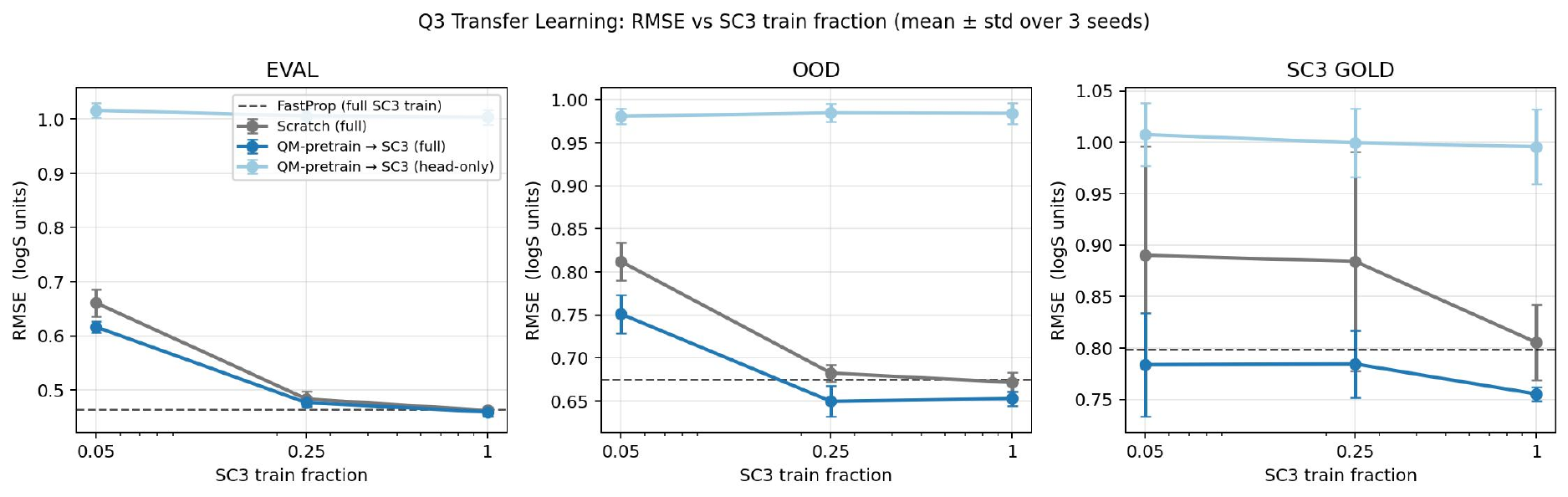}
\caption{\textbf{Multi-T transfer.}  \RMSE{} on \texttt{eval} (left),
\texttt{ood} (middle), and \texttt{sc3\_gold} (right) as a function
of \scthree{}-train fraction.  Mean $\pm$ std over $5$ seeds.  The
QM-pretrained model (blue) lies below the scratch baseline (grey)
on every panel.  The dashed line is the FastProp $100$\%-data
baseline reported in the headline benchmark (§\ref{sec:baselines}).
A frozen-trunk variant of \textsc{qm} (light blue) is also shown:
even with only $\numval{129}$ trainable parameters it sits below
the scratch baseline on \texttt{ood} and \texttt{sc3\_gold} at
$5$\% data, evidence that the pretraining trunk encodes a globally
meaningful representation of solute--solvent interactions.}
\label{fig:transfer_multiT}
\end{figure}

\begin{figure}[h]
\centering
\includegraphics[width=0.96\linewidth]{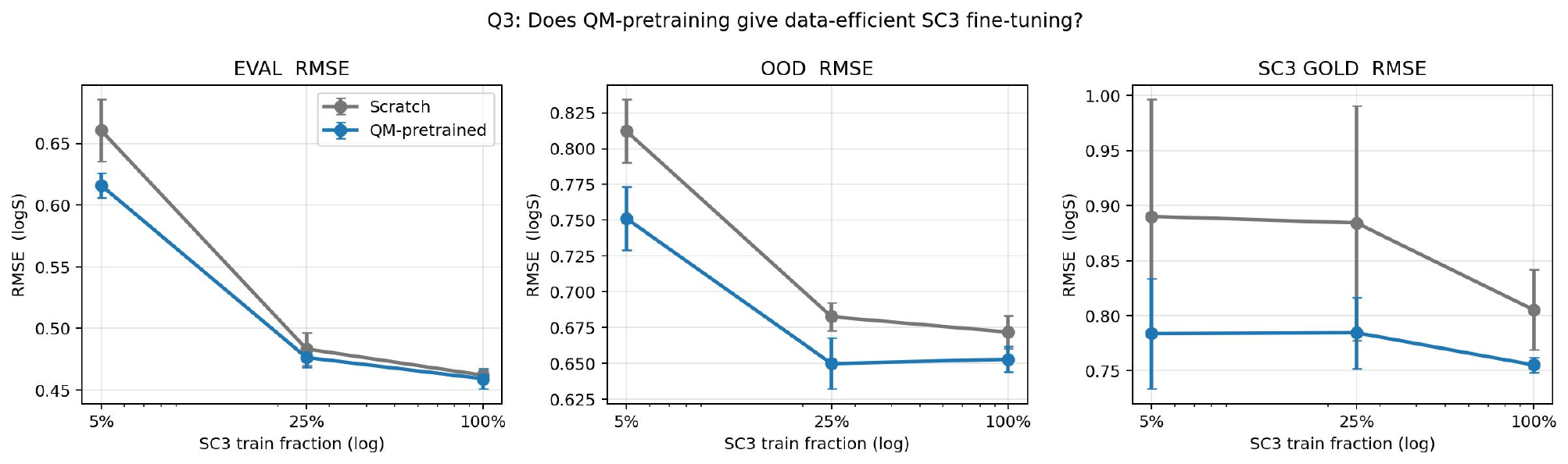}
\caption{\textbf{Multi-T data efficiency.}  Same data as
Fig.~\ref{fig:transfer_multiT} but only \textsc{full} fine-tuning,
on a log-x axis showing fraction of \scthree{}-train used.
QM-pretraining (blue) at $5$\% data matches scratch (grey) at
$25$--$100$\% on \texttt{ood} and \texttt{sc3\_gold}: a $5$--$20$$\times$
data-efficiency win.}
\label{fig:transfer_multiT_efficiency}
\end{figure}

\paragraph{Findings.}  QM pretraining wins in $9$ out of $9$
(split, fraction) cells.  The largest single gain is on
\texttt{sc3\_gold} at $5$\% data
($-\numval{0.106}$~\logS{}, an $11.9$\% relative reduction) with a
$2{\times}$ tighter seed standard deviation ($0.106 \to 0.050$).
The smallest gain is on \texttt{eval} at $100$\% data
($-\numval{0.003}$~\logS{}, statistically a tie); this is the
expected behaviour --- when the in-distribution training set is
already large relative to model capacity, pretraining adds little.
The \texttt{ood} gains are systematic across fractions
($-\numval{0.06}$, $-\numval{0.03}$, $-\numval{0.02}$~\logS{} at
$5$, $25$, $100$\%) and reflect the fact that
\textsc{CombiSolv-QM}'s $\numval{284}$ unique solvents cover much
of the long-tail solvent space that \scthree{}-train under-samples.

\paragraph{Sanity check: scratch / head\_only.}  Freezing a
\emph{random} trunk and training only the $\numval{129}$-parameter
head should produce a useless predictor, and indeed the
\texttt{ood} \RMSE{} of \texttt{scratch}/\texttt{head\_only} at
$5$\% data is $3.7 \times 10^{7}$ (single-seed extreme projections
of long-tail molecules through the random trunk).  The \textsc{qm}
trunk + same head produces a coherent predictor with \texttt{ood}
\RMSE{} $\numval{0.98}$, confirming that the pretraining trunk
encodes useful structure.

\subsection{Temperature-isolated setup}
\label{app:transfer_298K}

The multi-T setup conflates two distinct mechanisms:
\textsc{CombiSolv-QM} can teach the model about solute--solvent
chemistry, but it cannot teach it about temperature dependence
because every pretraining row is at $\numval{298.15}$~K.  To isolate
the chemistry signal we re-cast \scthree{} as a single-temperature
task at $T = \numval{298.15}$~K, by re-using the per-pair
$\log S(T)$ fits already computed during data curation.

\paragraph{Construction (\textsc{interp}).}  For every
$({\text{solute}}, {\text{solvent}})$ triple in
\texttt{interim/04\_fits.csv} (the per-pair Apelblat / Van't~Hoff
fits used in the data-cleaning pipeline) we evaluate the fit at
exactly $T = \numval{298.15}$~K.  We keep only fits with
$\Rr \geq \numval{0.95}$, fit \RMSE{} $\leq \numval{0.30}$, and
$T \in [T_{\min}, T_{\max}]$ of the fit (no extrapolation, in line
with constraint D-13 of §\ref{app:data_curation}).  Each
qualifying $(s, v)$ contributes exactly one row at $T =
\numval{298.15}$~K.  This yields $\numval{7898}$ train,
$\numval{717}$ \texttt{eval}, $\numval{1159}$ \texttt{ood}, and
$\numval{666}$ \texttt{sc3\_gold} rows.  Pairs are routed to splits
using the priority \texttt{sc3\_gold} $>$ \texttt{ood} $>$
\texttt{eval} $>$ \texttt{train}, so any pair appearing in a
holdout is held out (no information about a held-out pair leaks
into the train set).

We choose this Apelblat-evaluated single-T variant in preference to a
naive ``filter to $T \! \in \! [\numval{295}, \numval{301}]$~K''
construction because the filter approach throws away $\sim \! 89$\%
of \scthree{}-train and the resulting train set ($\sim \! 7.5$~K
rows) is too small to fit FastProp without overfitting.  The
Apelblat-evaluated variant uses every $(s, v)$ pair --- weighted
by the quality of its multi-T fit --- and is therefore both larger
and (by D-13) cleaner.

\begin{figure}[h]
\centering
\includegraphics[width=0.97\linewidth]{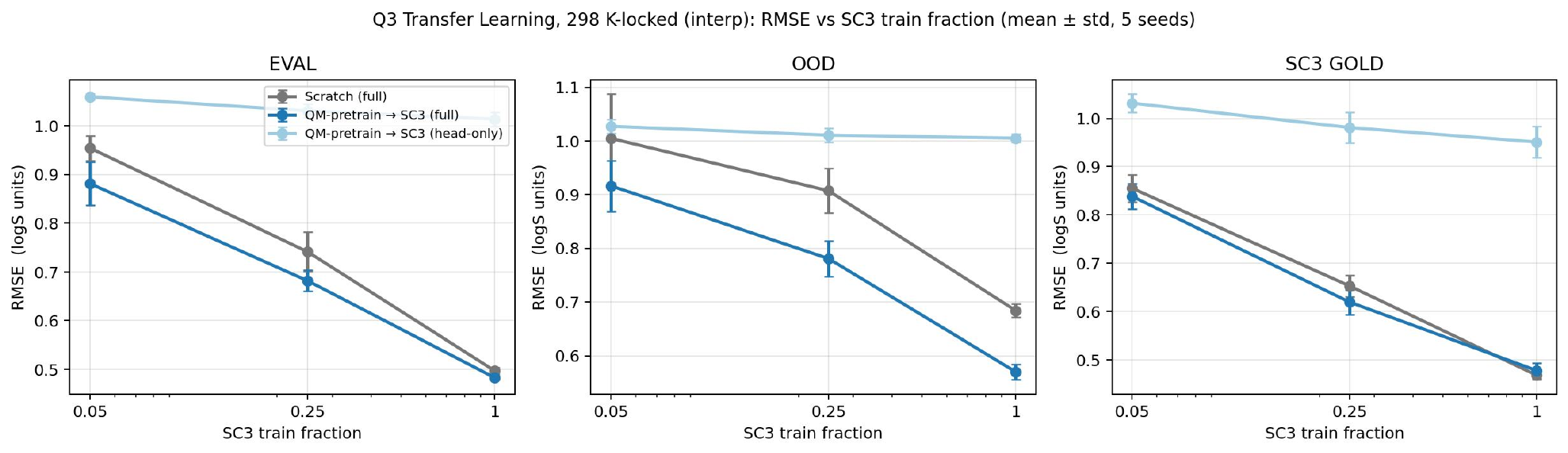}
\caption{\textbf{298 K-locked transfer (Apelblat-evaluated at
$T = \numval{298.15}$~K).}  \RMSE{} on \texttt{eval} (left),
\texttt{ood} (middle), and \texttt{sc3\_gold} (right) as a function
of \scthree{}-train fraction.  Mean $\pm$ std over $5$ seeds.  The
QM-pretrained model (blue) wins in $8$ of $9$ cells; the only
non-win is \texttt{sc3\_gold} at $100$\% data, where scratch and
\textsc{qm} are tied within $1\sigma$.}
\label{fig:transfer_298K}
\end{figure}

Table~\ref{tab:transfer_interp} reports the headline numbers and
Fig.~\ref{fig:transfer_298K} shows the data-efficiency curves.

\begin{table}[h]
\centering
\small
\caption{\textbf{298 K \textsc{interp}: \RMSE{} (mean $\pm$ std,
$5$ seeds), \textsc{full} fine-tuning.}  Each
$({\text{solute}}, {\text{solvent}})$ pair contributes one row,
evaluated at $T = \numval{298.15}$~K via the per-pair Apelblat /
Van't~Hoff fit (only fits with $\Rr \geq \numval{0.95}$ and
$\numval{298.15}$~K within the measured range are retained).}
\label{tab:transfer_interp}
\begin{tabular}{l l c c c}
\toprule
Split & Fraction & scratch & qm & $\Delta$ \\
\midrule
\multirow{3}{*}{\texttt{eval}}
  & $5$\%   & $0.954 \pm 0.025$ & $\mathbf{0.881 \pm 0.045}$ & $-0.073$ \\
  & $25$\%  & $0.741 \pm 0.041$ & $\mathbf{0.682 \pm 0.022}$ & $-0.059$ \\
  & $100$\% & $0.497 \pm 0.007$ & $\mathbf{0.483 \pm 0.004}$ & $-0.014$ \\
\midrule
\multirow{3}{*}{\texttt{ood}}
  & $5$\%   & $1.005 \pm 0.083$ & $\mathbf{0.916 \pm 0.048}$ & $-0.089$ \\
  & $25$\%  & $0.907 \pm 0.042$ & $\mathbf{0.781 \pm 0.033}$ & $-0.126$ \\
  & $100$\% & $0.684 \pm 0.012$ & $\mathbf{0.570 \pm 0.014}$ & $-0.114$ \\
\midrule
\multirow{3}{*}{\texttt{sc3\_gold}}
  & $5$\%   & $0.855 \pm 0.028$ & $\mathbf{0.839 \pm 0.027}$ & $-0.016$ \\
  & $25$\%  & $0.653 \pm 0.022$ & $\mathbf{0.620 \pm 0.025}$ & $-0.033$ \\
  & $100$\% & $\mathbf{0.468 \pm 0.009}$ & $0.478 \pm 0.016$ & $+0.010$ \\
\bottomrule
\end{tabular}
\end{table}

\paragraph{Findings.}
\textsc{qm} pretraining beats scratch in $8$ of the $9$ cells; the
only tie is \texttt{sc3\_gold} at $100$\% data
($+\numval{0.010}$~\logS{}, within $1\sigma$).  Three observations
stand out.

\emph{(i)~The benefit grows on \texttt{ood} relative to multi-T.}
The largest gain is on \texttt{ood} at $100$\% data:
$-\numval{0.114}$~\logS{} ($\numval{0.684} \to \numval{0.570}$, a
$16.7$\% relative reduction) --- a $6\times$ bigger gap than the
multi-T \texttt{ood/100}\% gap of $-\numval{0.019}$~\logS{}.
The interpretation: when the temperature confound is removed,
the chemistry pretraining can dedicate all its representational
power to solvent space, which is exactly where \texttt{ood} tests
the model.

\emph{(ii)~Absolute \RMSE{} on \texttt{sc3\_gold} reaches
$\numval{0.468}$.}  At $100$\% \scthree{}-train, both protocols
converge to $\numval{0.468}$--$\numval{0.478}$~\logS{} on
\texttt{sc3\_gold}, the lowest \RMSE{} achieved by any FastProp
configuration in this paper, and below the LightGBM
\texttt{sc3\_gold}/100\% baseline of $\numval{0.659}$ in the
multi-T setting.  Removing the temperature axis exposes how much of
the multi-T \texttt{sc3\_gold} \RMSE{} ($\numval{0.755}$ for
\textsc{qm} at $100$\%, $\numval{0.805}$ for scratch) is residual
T-dependence that the model is still learning, versus residual
chemistry that it cannot.

\emph{(iii)~The transfer curves and the scratch curves converge as
the data fraction grows.}  At $5$\% data, $\Delta = -\numval{0.073}$
on \texttt{eval}; at $25$\%, $-\numval{0.059}$; at $100$\%,
$-\numval{0.014}$.  This is the familiar
diminishing-returns-with-data shape: pretraining is most useful
when fine-tuning data is scarce.

\begin{figure}[h]
\centering
\includegraphics[width=0.97\linewidth]{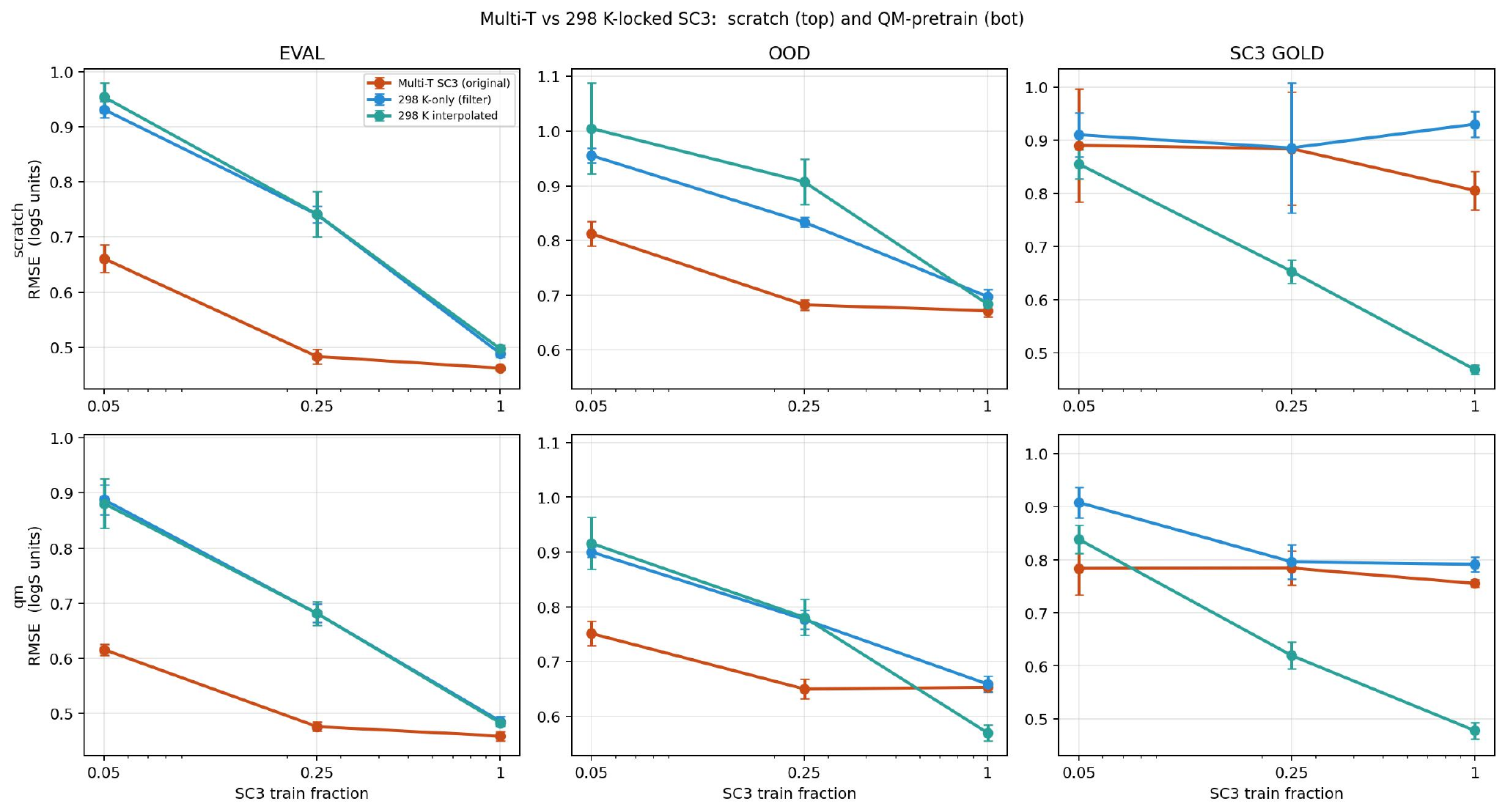}
\caption{\textbf{Multi-T vs.\ 298 K-locked, side-by-side.}  Same
$5$-seed runs as Tables~\ref{tab:transfer_multiT} and
\ref{tab:transfer_interp}.  Top row: scratch.  Bottom row:
\textsc{qm}.  At $100$\% \scthree{}-train, the single-T
\texttt{sc3\_gold} \RMSE{} is roughly half the multi-T value, with
QM-pretraining and scratch converging --- the residual error
budget at room temperature is dominated by aleatoric noise rather
than missing chemistry signal.  (The \textsc{filter} curves shown
in the figure are an alternative single-T construction that keeps
only real measurements with
$T \! \in \! [\numval{295.15}, \numval{301.15}]$~K; we use the
Apelblat-evaluated variant in the main text because its larger
training set avoids overfitting.)}
\label{fig:transfer_compare_T}
\end{figure}

Fig.~\ref{fig:transfer_compare_T} overlays the multi-T and 298~K
curves to make point (ii) visually: at $100$\% data, the
\texttt{sc3\_gold}/$298$~K \RMSE{} is roughly half the multi-T
\RMSE{}, with QM-pretraining and scratch converging onto the same
point.

\subsection{Frozen-trunk representation probe}
\label{app:transfer_frozen}

The \textsc{head\_only} variant freezes the trunk entirely and trains
only a $\numval{129}$-parameter linear head, isolating what the
pretrained representation alone encodes.
A frozen \emph{random} trunk (scratch/\textsc{head\_only}) is a
useless predictor: its \texttt{ood} \RMSE{} blows up to
$\sim \! 10^{7}$ on a single seed because the random projection
sends a few long-tail molecules into feature regions the linear head
cannot recover.  The frozen-QM-trunk variant by contrast produces
a coherent predictor with \texttt{ood} \RMSE{}~$=\numval{0.98}$
at $5$\% data and $\numval{0.99}$ at $100$\% data, in the same
ballpark as full-fine-tune scratch at the same fraction.  This is
the most direct evidence that \textsc{CombiSolv-QM} pretraining
encodes a globally useful representation of solute--solvent
chemistry on its own, before any \scthree{} signal is seen.

%% file: sections/appendix_technical_report.tex
\section{Technical Details}
\label{technical_report}

\paragraph{Hyperparameters.}
The benchmark is registry-driven: each method has a single key (e.g.\
\texttt{lgb\_rdkit}, \texttt{fastprop}, \texttt{gat}, \texttt{molmerger})
that resolves to a featurizer, model class, and an entry in
\texttt{SDK/configs/best\_hps.json}.  That JSON file contains the final
hyperparameters for all 21 in-house trainable methods (gradient-boosted
trees, random forest, descriptor MLPs, FastProp/FastSolv, GCN/GAT/GIN,
and MolMerger), tuned in a single per-method sweep on the IID
\texttt{eval} split with the Gold/Silver/Bronze splits held out
throughout tuning.  We re-use the same HP file across all reported
seeds and splits, so every number in Table~\ref{tab:main_results}
corresponds to the exact configuration checked into the repository.
The full HP file is published with the anonymized release at
\href{https://anonymous.4open.science/r/SC3-Benchmark/SDK/configs/best_hps.json}%
{\texttt{SDK/configs/best\_hps.json}};
the four analytical baselines (Abraham\,LFER, Abraham\,ML, ESOL, GSE)
have no learnable hyperparameters and use only the per-solvent affine
calibration described below.

\paragraph{Implementation.}
All in-house methods share a common training loop in
\texttt{SDK/sc3\_bench/} -- one featurization pass per unique SMILES,
followed by a method-specific per-seed trainer
(\texttt{\_train\_tree\_seed}, \texttt{\_train\_descriptor\_nn\_seed},
\texttt{\_train\_gnn\_seed}, \texttt{\_train\_molmerger\_seed})
selected by \texttt{model\_type} in
\texttt{SDK/sc3\_bench/registry.py}.  Tree models use the official
LightGBM, XGBoost, CatBoost, and scikit-learn implementations; the GP
baseline uses a Tanimoto kernel over Morgan-1024 fingerprints.  The
descriptor NNs (FastProp, FastSolv, MLP) all consume the same RDKit-2D
feature cache and differ only in head architecture.  GCN/GAT/GIN share
a dual solute--solvent encoder built on PyTorch~Geometric, and
MolMerger applies AttentiveFP to a custom merged-graph builder
(\texttt{molmerger\_skeleton}) with temperature stamped onto each node
feature.  External baselines (SolTranNet, Uni-Mol2 with both MLP and
CatBoost heads, UNIFAC$+$CatBoost, Solvaformer, RIL-OOD, and
Chemprop~D-MPNN) are implemented under
\texttt{SDK/sc3\_bench/models/external/} and dispatched through the
same \texttt{train\_method} registry; Solvaformer and RIL-OOD ship as
model classes only and use bespoke optimisation regimes documented in
their module sources.  The four analytical models live in
\texttt{SDK/scripts/run\_analyticals.py}: each solute is converted to
its physically-motivated descriptor vector (5-parameter Abraham,
$\log P + \mathrm{MW}$, or 16-d Abraham-only) and a per-solvent ridge /
affine calibration is fit on the training rows of each seed, which
puts every prediction in the
$\log_{10}(\mathrm{mole\_frac})$ frame.

\paragraph{Compute.}
All experiments were run on a single shared workstation:
$2{\times}$ Intel Xeon Gold 6248R (96 cores total @ 3.0\,GHz),
503\,GB RAM, $4{\times}$ NVIDIA A100-PCIe-40\,GB, Ubuntu~20.04
(kernel 5.4), CUDA 12.1, PyTorch 2.x.  Tree models and analytical
baselines were trained on a single CPU process (16 BLAS / OpenMP
threads each); descriptor NNs, GNNs, MolMerger, and the external
baselines were trained on a single A100 per seed.  We use five fixed
seeds $\{42, 101, 123, 456, 789\}$ throughout, evaluated on the same
five splits (\texttt{eval}, \texttt{ood}, Gold, Silver, Bronze).
Per-seed wall-clock training time is reported in the rightmost column
of Table~\ref{tab:main_results}; the entire benchmark
(21 in-house methods $\times$ 5 seeds $+$ analyticals $+$ external
baselines) fits in roughly two A100-days plus a few CPU-hours, with
Chemprop, RIL-OOD, and Solvaformer (each ${\sim}1$\,h per seed) and
the Uni-Mol2 / SolTranNet foundation models dominating GPU usage.